\documentclass[journal]{IEEEtran}
\usepackage{latexsym,bm}
\usepackage[tbtags]{amsmath}
\usepackage[dvips]{graphicx,color}
\usepackage{subfigure}
\usepackage{amssymb}
\usepackage{cite}
\usepackage{stfloats}

\usepackage{times}
\usepackage{caption}
\usepackage{amsmath}
\usepackage{amssymb}
\usepackage{amscd}
\usepackage{cite}
\usepackage{amsfonts}
\usepackage{graphicx}
\usepackage{psfrag}
\usepackage{epsfig}
\usepackage{algorithm}
\usepackage{algorithmic}
\ifCLASSINFOpdf
\else
\fi
\begin{document}
%
\title{Destination-aided Wireless Power Transfer in Energy-limited Cognitive Relay Systems}
\author{ Ruijin Sun, Ying Wang,  Zhongyu Miao, and Xinshui Wang
\thanks{This paragraph of the first footnote will contain the  date on which you submitted your paper for review. This paragraph of the first footnote will contain the  date on which you submitted your paper for review. This work was supported by National Natural Science Foundation of China (Project 61431003, 61421061) and National 863 Project 2014AA01A705. }
\thanks{R. Sun, Y. Wang,  Z. Miao, and X. Wang are with the State Key Laboratory of Networking and Switching Technology, Beijing University of Posts and Telecommunications, Beijing 100876, P.R. China
(email:  sunruijin1992@gmail.com, wangying@bupt.edu.cn, zhongyumiao@139.com, wxinshui@126.com). }
\thanks{Corresponding author: Y. Wang (wangying@bupt.edu.cn) }}
\maketitle

\begin{abstract}
   This paper considers an energy-limited  cognitive relay network where a secondary transmitter (ST) assists to forward the traffic from a primary transmitter (PT) to a primary receiver (PR), in exchange for serving its own secondary receiver (SR) in the same frequency. The multiple-antenna ST is assumed to be energy-constrained  and  powered by both information flow from source (PT) and dedicated energy streams from destinations (PR and SR),  which is called   destination-aided wireless power transfer (DWPT) scheme. Then, the relay processing matrix, cognitive beamforming vector and power splitter are jointly designed to maximize the rate of secondary users  under the energy causality constraint and the constraint that the demanded rate of primary users is satisfied.  For the perfect channel information state (CSI) case, by adopting semi-definite relax (SDR) technique and  Charnes-Cooper transformation, the global optimal solution is given. To reduce the complexity, matrix decomposition, zero forcing (ZF) scheme, and dual method are jointly employed to derive a suboptimal solution. For the imperfect CSI case, S-procedure is used to transform the worst-case robust problem into a tractable semi-definite program (SDP). Simulation results reveal that our proposed  DWPT scheme is greatly preferred for both perfect and imperfect CSI cases when ST is close to PR/SR.
\end{abstract}


%
\IEEEpeerreviewmaketitle

\section{Introduction}
%
%
%
%

\IEEEPARstart{W}{ireless} power transfer,  potentially  enabling low-power cost systems to work self-sustainably, has attracted considerable attention recently. Since  electromagnetic waves can carry both energy and information, simultaneous wireless information and power transfer (SWIPT) is first creatively proposed in \cite{varshney2008transporting}. In that paper, the same received radio frequency (RF) flow is ideally assumed to be extracted for both information decoding (ID) and  energy harvesting (EH), which cannot be conducted in recent electric circuits. Then, two basic practical receiver architectures for SWIPT named time switching (TS) and power splitting (PS) are put forward \cite{zhang2013mimo}.
TS switches the receiver between ID and EH in a time-division manner, while PS   divides the received stream into two flows with one for ID and the other for EH.
Since then, several techniques in wireless communications  have been extended to SWIPT systems, including multiple-input  multiple-output (MIMO) \cite{zhang2013mimo,zong2016optimal}, relay \cite{huang2015joint,zhao2015joint,chen2015joint,nasir2014throughput}, cognitive radio \cite{zheng2014information,wang2015optimization}, and full duplex \cite{wang2016transceiver,wang2016beamforming}.
 In the SWIPT relay systems, energy-limited relays are able to  assist the traffic from sources to destinations with the scavenged energy from sources as their transmission power. Based on the proposed TS or PS scheme,    resource allocation and beamforming design \cite{zhao2015joint,chen2015joint, huang2015joint} in relay systems are  widely studied.

In  cognitive relay networks, secondary users (SU) are encouraged to relay primary users' (PU) messages for accessing the licensed spectrum to send their own information to secondary receivers (SR). It is a win-win strategy especially when the direct links between primary transmitters (PT) and primary receivers (PR) suffer from severe fading. With the EH ability, energy-limited  secondary transmitters (ST) can be strongly stimulated by both the information and  energy cooperation from PT to ST, which is investigated in \cite{zheng2014information}. In that paper, the ST is powered by PT first, and then uses the harvested energy as transmission power to forward PT's information to PR as well as to send its own information to SR. The SU rate maximization problem is considered subject to the PU rate demand constraint and the energy causality constraint. It has been found that, the SU-PU rate region can be  enlarged with the energy cooperation from PT to ST.

However, for the uplink transmission in sensor networks or other low-power networks, the power budget of PT is strictly restricted by its battery capacity,  while the destinations  are information access points, which usually have constant power supply. To better explore the system  performance for this practical scenario, power of PR and SR is fully exploited in this paper. In particular, a destination-aided wireless power transfer (DWPT) scheme is proposed for a cognitive relay system, where the energy-limited ST is not only powered by PT, but also assisted by energy transfer from PR and SR.

In \cite{huang2015joint}, a relay node powered  by both source and destination is investigated for a simple three-node SWIPT relay system. The system rate maximization problem is studied with the energy causality constraint. It is worth pointing out that, our considered DWPT scheme for cognitive relay system is a more general scenario as compared with \cite{huang2015joint}. If ST is a pure relay and does not send message to SR (of course, the rational SR  also does not  transfer energy to ST), our considered scenario will degrade into a three-node relay system in \cite{huang2015joint}. If ST is a pure transmitter and does not forward  the traffic from PT to PR (in this case, PR also dose not transfer energy to ST), our considered scenario will become a two-node wireless powered communication network in \cite{ju2014throughput}, where the downlink energy transfer and
uplink information transmission are assumed.

 To be specific, this paper studies a cognitive relay network where a PT, a PR, an energy-limited ST and a SR   are included. The ST is assumed to have multiple antennas and other users all have a single antenna. In the first phase, PT transmits information flow to ST. At the same time, PR and SR also send dedicated wireless RF energy stream to ST to further enhance the scavenged  energy at ST. Based on the PS scheme, the received RF flow at ST can be split  for EH and ID.  In the second phase, ST assists to relay the traffic from PT to PR with the amplify-and-forward (AF) protocol and also sends its own information to SR.  The main contributions of this work are listed as follows.

 1)  A DWPT scheme for the cognitive relay system is first proposed and investigated, which fully exploits the power of destinations. With this scheme, ST can extract energy from both information flow from PT and dedicated energy flows from destinations (PR and SR). The relay processing matrix, cognitive beamforming vector and power splitter are jointly optimized to maximize the SU rate under the energy causality constraint and the constraint that   minimal PU rate demand is guaranteed.

 2) Under the assumption that ST perfectly knows all the channel state information (CSI), both the optimal and low-complexity suboptimal solutions are given. Since the problem is non-convex,  iterative approaches are presented. With given power splitter, to achieve the optimal relay matrix and cognitive beamforming vector, we first derive  lower dimensional structures for them, and then adopt the Charnes-Cooper transformation and semi-definite relax (SDR) method. To reduce the complexity, matrix decomposition, zero forcing (ZF) scheme, and dual method are jointly employed to derive the closed-form solution. Then, the optimal power splitters for both optimal and suboptimal algorithms are found via bisection.

3) Under the assumption that ST imperfectly knows the channel links from ST to PR and  ST to SR, a  worst-case robust solution is proposed in an iterative manner. With fixed power splitter, to find the relay matrix and cognitive beamforming vector, some matrix lemmas and S-procedure are used to transform the robust problem into a tractable semi-definite program (SDP). Then, the optimal power splitter is found via one-dimensional search.

4) For comparison, the energy harvesting cognitive radio system without destinations' power transfer in \cite{zheng2014information} is also considered. Simulation results reveal that, when ST is close to PR/SR, our proposed DWPT scheme  is greatly preferred for both perfect and imperfect CSI cases.

The remainder of the paper is organized as follows.  In Section \uppercase\expandafter{\romannumeral2}, system model and problem formulation are introduced. In Section \uppercase\expandafter{\romannumeral3}, we present both optimal and suboptimal solutions to the SU rate maximization problem with the perfect CSI.  In Section \uppercase\expandafter{\romannumeral4}, we further state a worst-case robust algorithm for the problem with the imperfect CSI. The simulation results are presented and discussed in Section \uppercase\expandafter{\romannumeral5}. Finally, Section \uppercase\expandafter{\romannumeral6} concludes the paper.

\emph{Notation}:  Bold lower and upper case letters are used to denote column vectors and matrices, respectively. The superscripts ${{\mathbf{H}}^T}$, ${{\mathbf{H}}^*}$ and ${{\mathbf{H}}^H}$ is  standard transpose,  conjugate and (Hermitian) conjugate transpose  of $\mathbf{H}$, respectively. $\left\| {\mathbf{.}}\right\|_2$ and $\left\| {\mathbf{.}} \right\|_F$ refer to the Euclidean norm  and the Frobenius norm, respectively. $\operatorname{Rank}(\mathbf{W})$, $\operatorname{Tr}(\mathbf{W})$ and $\text{vec}(\mathbf{W})$ denote the rank, trace and  vectorization of matrix $\mathbf{W}$, respectively. $\mathbf{W} \succeq \mathbf{0} (\preceq \mathbf0)$ means that matrix $\mathbf{W}$ is positive semidefinite (negative semidefinite).  $\otimes$ and $\odot$
are Kronecker product and Hadamard product, respectively. Matrix $\mathbf{E}$ represents $\text{diag}(\mathbf{1}, \mathbf{1},..., \mathbf{1})$.



%

\section{System Model and Problem Formulation}

Considering a  cognitive relay network where a ST assists to forward the traffic from a PT to a  PR, in exchange for serving its own SR in the same frequency, as illustrated in Fig. 1. The ST is equipped with $M$ antennas while other users have a single antenna. We assume that the ST is energy-limited, and thus  powered by PT as well as PR and SR to enhance the harvested energy,  which is called DWPT scheme in this paper.
Assume that  the entire communication time slot, which consists of two equal phases, is normalized to be 1.

\begin{figure}
\centering
\includegraphics[width=6.2cm, height=3.8cm]{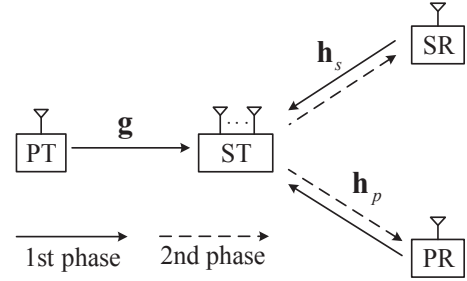}
\caption{Two-phase DWPT scheme in an energy-limited four-node cognitive relay network.}
\label{systemmodel}
\end{figure}

In the first phase, PT transmits information signal $x_p$ with power $P_{PT}$. Meanwhile, PR and SR respectively send energy signals $x_p'$ and $x_s'$ with their corresponding power, $P_{PR}$ and $P_{SR}$. Suppose that different signals are statistically independent. The observation at ST is expressed as
\begin{equation}
{{\mathbf{y}}_r} = {\mathbf{g}}{x_p} + {{\mathbf{h}}_p}{x'_p} + {{\mathbf{h}}_s}{x'_s} + {{\mathbf{n}}_r},
\end{equation}
where $\mathbf{g}$, $\mathbf{h}_p$ and $\mathbf{h}_s$ are $M \times 1$ channel vectors from PT to ST, PR to ST, and SR to ST, respectively; ${{\mathbf{n}}_r}\sim {\cal CN}(0,\sigma _r^2{{\mathbf{I}}_M})$ is the received noise vector at ST. To concurrently process the information decoding and the energy harvesting, the practical PS receiver architecture is adopted at ST. In particular, the received signal is split into two streams, one for ID and one for EH, with the relative power ratio of $\rho$ and $1-\rho$, respectively. The signal stream for EH and the harvested energy at ST are respectively given by
\begin{equation}
{\mathbf{y}}_r^{EH} = \sqrt {1 - \rho } \left( {{\mathbf{g}}{x_p} + {{\mathbf{h}}_p}{{x}_p'} + {{\mathbf{h}}_s}{{x}_s'} + {{\mathbf{n}}_r}} \right)~\text{and}
\end{equation}
\begin{align}
P_r^{EH} \!= \!\xi \left( {1 - \rho } \right)\!( {P_{PT}}\left\| {\mathbf{g}} \right\|_2^2 \!+\!  {P_{PR}}\left\| {{{\mathbf{h}}_p}} \right\|_2^2 \!+\! {P_{SR}}\left\| {{{\mathbf{h}}_s}} \right\|_2^2 \!+ \!\sigma _r^2 ),
\end{align}
where $0<\xi \leq 1$ is the energy conversion efficiency.
Let ${{\mathbf{n}}_c}\sim {\cal CN}(0,\sigma _c^2{{\mathbf{I}}_M})$
  denote the $N \times 1$ circuit noise vector caused by the signal frequency conversion from RF to baseband and hence the other stream for ID is given by
 \begin{equation}
 {\mathbf{y}}_r^{ID} = \sqrt \rho  \left( {{\mathbf{g}}{x_p} + {{\mathbf{h}}_p}{{x}_p'} + {{\mathbf{h}}_s}{{x}_s'} + {{\mathbf{n}}_r}} \right) + {{\mathbf{n}}_c}.
 \end{equation}

During the second phase, ST uses the AF protocol to relay the traffic from PT to PR and also superimposes its own message $x_s$ to SR with ${\rm E}\left[ {{{\left| {{x_s}} \right|}^2}} \right] = 1$. Denote  ${\mathbf{F}} \in {\mathbb{C}^{M \times M}}$ and  ${\mathbf{w}} \in {\mathbb{C}^{M \times 1}}$ as the relay processing matrix and the cognitive beamforming vector respectively, the transmit signal at ST is
\begin{align}
{{\mathbf{x}}_r} &= {\mathbf{Fy}}_r^{ID} + {\mathbf{w}}{x_s} \nonumber \\
&= {\mathbf{F}}\left( {\sqrt \rho  \left( {{\mathbf{g}}{x_p} + {{\mathbf{h}}_p}{{x}_p'} + {{\mathbf{h}}_s}{{x}_s'} + {{\mathbf{n}}_r}} \right) + {{\mathbf{n}}_c}} \right) + {\mathbf{w}}{x_s}
\end{align}
with average power
\begin{align}
{P_{ST}}({{\mathbf{F}},{\mathbf{w}},\rho}) =&  \rho \big({P_{PT}} \left\| {{\mathbf{Fg}}} \right\|_2^2 + {P_{PR}} \left\| {{\mathbf{F}}{{\mathbf{h}}_p}} \right\|_2^2 + {P_{SR}} \left\| {{\mathbf{F}}{{\mathbf{h}}_s}} \right\|_2^2 \nonumber \\
&+  \sigma _r^2\left\| {\mathbf{F}} \right\|_F^2 \big)
 + \sigma _c^2\left\| {\mathbf{F}} \right\|_F^2 + \left\| {\mathbf{w}} \right\|_2^2.
\end{align}

With the perfect CSI at ST, we assume that  PR/SR  can successfully cancel its  self-interference $x_p'$/$x_s'$.  Hence, the received signals at PR and SR are respectively expressed as
\begin{align}
{y_p} = & \sqrt \rho {\mathbf{h}}_p^H{\mathbf{Fg}}{x_p} +\sqrt \rho {\mathbf{h}}_p^H
{\mathbf{F}}{{\mathbf{h}}_s}{x'_s}+ {\mathbf{h}}_p^H{\mathbf{w}}{x_s} \nonumber \\
& +\sqrt \rho {\mathbf{h}}_p^H{\mathbf{F}}{{\mathbf{n}}_r}
+ {\mathbf{h}}_p^H {\mathbf{F}}{{\mathbf{n}}_c} + {n_p},
\end{align}
\begin{align}
{y_s} =&\sqrt \rho {\mathbf{h}}_s^H {\mathbf{Fg}}{x_p} +  \sqrt \rho {\mathbf{h}}_s^H  {\mathbf{F}}{{\mathbf{h}}_p}{x'_p} + \mathbf{h}_s^H{\mathbf{w}}{x_s}\nonumber\\
&+  \sqrt \rho {\mathbf{h}}_s^H {\mathbf{F}}{{\mathbf{n}}_r} + {\mathbf{h}}_s^H {\mathbf{F}}{{\mathbf{n}}_c} + {n_s},
\end{align}
where ${{{n}}_p}\sim {\cal CN}(0,\sigma _p^2)$ and ${{{n}}_s}\sim {\cal CN}(0,\sigma _s^2)$ are additive Gaussian white noises (AWGNs) at PR and SR, respectively.
  The received signal to interference plus noise  ratios (SINRs) at PR and SR are respectively given by
\begin{align}
&{\Gamma _p}({{\mathbf{F}}, {\mathbf{w}}, \rho}) = \\
&\frac{{\rho {P_{PT}}{{\left| {{\mathbf{h}}_p^H{\mathbf{Fg}}} \right|}^2}}}
{{\rho {P_{SR}}{{\left| {{\mathbf{h}}_p^H{\mathbf{F}}{{\mathbf{h}}_s}} \right|}^2} \!+\!(\rho \sigma _r^2+\sigma _c^2)\left\| {{\mathbf{h}}_p^H{\mathbf{F}}} \right\|_2^2 \!+\! \left| {{\mathbf{h}}_p^H{\mathbf{w}}} \right|^2 \!+\! \sigma _p^2}}, \nonumber
\end{align}
\begin{align}
&{\Gamma _s}({{\mathbf{F}}, {\mathbf{w}}, \rho}) = \\
&\frac{{\left| {{\mathbf{h}}_s^H{\mathbf{w}}} \right|^2}}
{{\rho( {P_{P\!T}}{{\left| {{\mathbf{h}}_s^H{\mathbf{Fg}}} \right|}^2}\! +\! {P_{P\!R}}{{\left| {{\mathbf{h}}_s^H{\mathbf{F}}{{\mathbf{h}}_p}} \right|}^2}) \!+\!(\rho \sigma _r^2\!+ \!\sigma _c^2)\!\left\| {{\mathbf{h}}_s^H{\mathbf{F}}} \right\|_2^2\! +\!  \sigma _s^2}}. \nonumber
\end{align}

In this paper, we focus on the joint design of relay matrix $\mathbf{F}$, cognitive beamforming vector $\mathbf{w}$ and  power splitter $\rho$ to maximize the achieved rate of SU, under the constraint that the rate demand of PU, ${R_p^{min}}$, is met. The optimization problem is formulated as ($\mathcal{P}1$)
\begin{subequations}
\begin{align}
&\underset{{{\mathbf{F}},{\mathbf{w}},0\leq\rho\leq 1}}{\text{max}}&& \!\!\!\!\!\!\!\! \Gamma_{s}( {{\mathbf{F}},{\mathbf{w}}, \rho} ) \\
&~~\text{s. t.}&& \!\!\!\!\!\!\!\!\!\!\!\!\!\!\!\!\Gamma_{p}( {{\mathbf{F}},{\mathbf{w}},\rho}) \geq \Gamma_p^{min}, \\
&&&\!\!\!\!\!\!\!\!\!\!\!\!\!\!\!\!\!\!\!\!\!\!\!\!P_{ST}({{\mathbf{F}},{\mathbf{w}},\rho}) \leq P_r^{EH},
\end{align}
\end{subequations}
where $\Gamma_p^{min}=2^{R_p^{min}}-1$.  Constraint (11c) is to guarantee that the transmission power at ST is not more than its harvested energy.

\section{ Solutions to DWPT with Perfect CSI}
In this section, we suppose that all the channel knowledge is perfectly known at ST. Under this assumption, both the optimal and low-complexity suboptimal solutions to problem $\mathcal{P}1$ are proposed.

\subsection{Optimal Solution to DWPT}
In this subsection, we present the optimal solution to problem $\mathcal{P}1$. Given fixed $\rho$,  optimal $\mathbf{F}$ and $\mathbf{w}$ are obtained, and then  the optimal $\rho$ is found via bisection. In what follows, we first  focus on the design of optimal $\mathbf{F}$ and $\mathbf{w}$ with fixed $\rho$.

\textit{\textbf{Proposition 1}:
Define two QR decompositions as  $\left[ {{{\mathbf{h}}_s},{\text{ }}{{\mathbf{h}}_p}} \right] = {{\mathbf{V}}_1}{{\mathbf{R}}_1}$ and $\left[ {{{\mathbf{h}}_s},{\text{ }}{{\mathbf{h}}_p},{\text{ }}{\mathbf{g}}} \right] = {{\mathbf{U}}_2}{{\mathbf{R}}_2}$, where ${{\mathbf{V}}_1} \in {\mathbb{C}^{M \times 2}}$, ${{\mathbf{U}}_2} \in {\mathbb{C}^{M \times 3}}$  are orthonormal matrices and   ${{\mathbf{R}}_1} \in {\mathbb{C}^{2 \times 2}}$, ${{\mathbf{R}}_2} \in {\mathbb{C}^{3 \times 3}}$ are upper triangular matrices. Then, the optimal relay matrix and cognitive beamforming vector have the following structures:
\begin{equation}
{\mathbf{F}}{\text{ }}={{\mathbf{V}}_1}{\mathbf{AU}}_2^H, ~~ {\mathbf{w}}{\text{ }}={{\mathbf{V}}_1}{\mathbf{b}},
 \end{equation}
 where ${\mathbf{A}} \in {\mathbb{C}^{2 \times 3}}$ and ${\mathbf{b}} \in {\mathbb{C}^{2 \times 1}}$
 are optimization variables.}

\textbf{Proof}: Please see Appendix A.
 \IEEEQED

If $M\geq3$, with \textbf{Proposition 1}, the $M^2$ unknowns in $\mathbf{F}$ and $M$ unknowns in $\mathbf{w}$ are respectively reduced to $2\times3$ unknowns and $2$ unknowns. This greatly reduces the computational complexity of the beamforming design. If $M\leq2$, we optimize $\mathbf{F}$ and $\mathbf{w}$ directly.

Define  ${{\mathbf{\hat h}}_s} = {\mathbf{V}}_1^H{{\mathbf{h}}_s}$, ${{\mathbf{\hat h}}_p} = {\mathbf{V}}_1^H{{\mathbf{h}}_p}$, ${{\mathbf{\bar h}}_s} = {\mathbf{U}}_2^H{{\mathbf{h}}_s}$, ${{\mathbf{\bar h}}_p} = {\mathbf{U}}_2^H{{\mathbf{h}}_p}$ and ${\mathbf{\bar g}} = {\mathbf{U}}_2^H{\mathbf{g}}$, with  \textbf{Proposition 1}, the  problem $\mathcal{P}1$  can be reformulated as (given fixed $\rho$) ($\mathcal{P}2$)
\begin{subequations}
\begin{align}
&\!\!\underset{{{\mathbf{A}},{\mathbf{b}}}}{\text{max}}&& \!\!\!\!\frac{{\left| {{\mathbf{\hat h}}_s^H{\mathbf{b}}} \right|^2}}
{{\rho \big({P_{P\!T}}{{\left| {{\mathbf{\hat h}}_s^H{\mathbf{A\bar g}}} \right|}^2} \!\!+ \!\! {P_{P\!R}}{{\left| {{\mathbf{\hat h}}_s^H{\mathbf{A}}{{{\mathbf{\bar h}}}_p}} \right|}^2}\big)\!\! +\!\! (\rho \sigma _r^2 \!\!+\!\! \sigma _c^2)\!\left\| {{\mathbf{\hat h}}_s^H{\mathbf{A}}} \right\|_2^2 \!\!+ \!\!\sigma _s^2}}  \\
&\!\!\text{s. t.}&&\!\!\!\!\!  \frac{{\rho {P_{PT}}{{\left| {{\mathbf{\hat h}}_p^H{\mathbf{A\bar g}}} \right|}^2}}}
{{\rho {P_{SR}}{{\left| {{\mathbf{\hat h}}_p^H{\mathbf{A}}{{{\mathbf{\bar h}}}_s}} \right|}^2} \!\!+ \!\!(\rho \sigma _r^2\!\! +\!\! \sigma _c^2)\!\left\| {{\mathbf{\hat h}}_p^H{\mathbf{A}}} \right\|_2^2 \!\!+\!\! \left| {{\mathbf{\hat h}}_p^H{\mathbf{b}}} \right|^2 \!\!+\!\! \sigma _p^2}}\!\! \geq\!\! \Gamma _p^{\min }, \displaybreak[0]\\
&&& \!\!\!\! \rho({P_{PT}} \left\| {{\mathbf{A\bar g}}} \right\|_2^2 + {P_{PR}} \left\| {{\mathbf{A}}{{{\mathbf{\bar h}}}_p}} \right\|_2^2 + {P_{SR}} \left\| {{\mathbf{A}}{{{\mathbf{\bar h}}}_s}} \right\|_2^2) \nonumber \displaybreak[0] \\
&&& + (\rho \sigma _r^2 + \sigma _c^2)\left\| {\mathbf{A}} \right\|_F^2 + \left\| {\mathbf{b}} \right\|_2^2 \leq P_r^{EH}.
\end{align}
\end{subequations}

To solve this problem effectively, we use equations
\begin{equation}
\operatorname{vec} ({\mathbf{AXB}}) = ({{\mathbf{B}}^T} \otimes {\mathbf{A}})\operatorname{vec} ({\mathbf{X}})~\text{and}
\end{equation}
\begin{equation}
{\mathbf{Tr}}({\mathbf{X}}_1^T{{\mathbf{X}}_2}) = {\mathbf{vec}}{({{\mathbf{X}}_1})^T}{\mathbf{vec}}({{\mathbf{X}}_2})
 \end{equation}
 to further rewritten problem $\mathcal{P}2$ as ($\mathcal{P}2.1$)
\begin{subequations}
\begin{align}
&\!\!\!\!\!\underset{{{\mathbf{a}},{\mathbf{b}}}}{\text{max}}&& \frac{{{{\mathbf{b}}^H}{{{\mathbf{\hat H}}}_s}{\mathbf{b}}}}
{{{{\mathbf{a}}^H}{{\mathbf{B}}_\rho }{\mathbf{a}} + \sigma _s^2}}  \\
&\!\!\!\!\!\text{s. t.}&&\!\!\!\!\!\!\! \frac{{{{\mathbf{a}}^H}{{\mathbf{C}}_\rho }{\mathbf{a}}}}{{{{\mathbf{a}}^H}{{\mathbf{D}}_\rho }{\mathbf{a}} + {{\mathbf{b}}^H}{{{\mathbf{\hat H}}}_p}{\mathbf{b}} + \sigma _p^2}} \geq \!\! \Gamma _p^{\min},\\
&&&{{\mathbf{a}}^H}{{\mathbf{E}}_\rho }{\mathbf{a}} +  {{\mathbf{b}}^H}{\mathbf{b}} \leq P_r^{EH},
\end{align}
\end{subequations}
where
\begin{equation}
{{\mathbf{B}}_\rho } = \rho \left( {\sigma _r^2{\mathbf{I}} + {P_{PT}}{{{\mathbf{\bar G}}}^T} + {P_{PR}}{\mathbf{\bar H}}_p^T} \right) \otimes {{{\mathbf{\hat H}}}_s} + \sigma _c^2{\mathbf{I}} \otimes {{{\mathbf{\hat H}}}_s},
\end{equation}
\begin{equation}
{{\mathbf{C}}_\rho } = \rho {P_{PT}}{{\mathbf{\bar G}}^T} \otimes {{\mathbf{\hat H}}_p},
\end{equation}
\begin{equation}
{{\mathbf{D}}_\rho } = \rho \left( {\sigma _r^2{\mathbf{I}} + {P_{SR}}{\mathbf{\bar H}}_s^T} \right) \otimes {{{\mathbf{\hat H}}}_p} + \sigma _c^2{\mathbf{I}} \otimes {{{\mathbf{\hat H}}}_p},
\end{equation}
\begin{align}
{{\mathbf{E}}_\rho } = &\rho \big( ({P_{PT}}{{{\mathbf{\bar G}}}^T} + {P_{PR}}{\mathbf{\bar H}}_p^T + {P_{SR}}{\mathbf{\bar H}}_s^T) \otimes {{\mathbf{I}}_{2 \times 2}} \\ &+ \sigma _r^2{{\mathbf{I}}_{6 \times 6}} \big)
+ \sigma _c^2{{\mathbf{I}}_{6 \times 6}}, ~\text{and} \nonumber
\end{align}
\begin{equation}
{\mathbf{a}} = \operatorname{vec} ({\mathbf{A}}), {{\mathbf{\hat H}}_s} =  {{\mathbf{\hat h}}_s}{\mathbf{\hat h}}_s^H, {{\mathbf{\hat H}}_p} = {{\mathbf{\hat h}}_p}{\mathbf{\hat h}}_p^H,
\end{equation}
\begin{equation}
{{\mathbf{\bar H}}_s} = {{\mathbf{\bar h}}_s}{\mathbf{\bar h}}_s^H, {{\mathbf{\bar H}}_p} = {{\mathbf{\bar h}}_p}{\mathbf{\bar h}}_p^H, {\mathbf{\bar G}} = {\mathbf{\bar g}}{{\mathbf{\bar g}}^H}.
\end{equation}

 Before solving problem $\mathcal{P}$2.1,  we first analyze its feasible condition, which can be obtained by finding the maximum $\Gamma_p^{min*}$. Setting $\mathbf{b}=\mathbf{0}$, the optimization problem is given as ($\mathcal{P}2.2$)
  \begin{subequations}
\begin{align}
&\underset{{{\mathbf{a}}}}{\text{max}}&& \!\!\!\!\!\!\!\!\!\!\!\!\!\!\!\!\!\!\!\!\!\!\!\!\!\!\!\!\!\!\!\!\!\!\! \frac{{{{\mathbf{a}}^H}{{\mathbf{C}}_\rho }{\mathbf{a}}}}{{{{\mathbf{a}}^H}{{\mathbf{D}}_\rho }{\mathbf{a}}  + \sigma _p^2}}  \\
&\text{s. t.}&&\!\!\!\!\!\!\!\!\!\!\!\!\!\!\!\!\!\!\!\!\! \!\!\!\!\!\!\!\!\!\!\!\!\!\! {{\mathbf{a}}^H}{{\mathbf{E}}_\rho }{\mathbf{a}} \leq P_r^{EH}.
\end{align}
\end{subequations}
 It is easy to verify that, at the optimum, the power constraint (23b) is active, i.e., ${{\mathbf{a}}^H}{{\mathbf{E}}_\rho }{\mathbf{a}} = P_r^{EH}$. With this equation, problem $\mathcal{P}2.2$ can be equivalently written as ($\mathcal{P}$2.3)
 \begin{equation}
\underset{{{\left\| {\mathbf{a}} \right\|_2^2} = 1}}{\text{max}}~~ \frac{{{{\mathbf{a}}^H}{{\mathbf{C}}_\rho }{\mathbf{a}}}}{{{{\mathbf{a}}^H}\big({{\mathbf{D}}_\rho }+  {{\sigma _p^2} \mathord{\left/
 {\vphantom {{\sigma _p^2} {P_r^{EH}}}} \right.
 \kern-\nulldelimiterspace} {P_r^{EH}}}{{\mathbf{E}}_\rho } \big){\mathbf{a}} }}
\end{equation}
which is a generalized Rayleigh quotient \cite{horn2012matrix}. The optimal $\mathbf{a}^*$ of problem $\mathcal{P}2.3$ is equal to the dominant generalized eigenvector of the matrix pair $\big({{\mathbf{C}}_\rho },~ {{\mathbf{D}}_\rho }+  {{\sigma _p^2} \mathord{\left/ {\vphantom {{\sigma _p^2} {P_r^{EH}}}} \right.
 \kern-\nulldelimiterspace} {P_r^{EH}}}{{\mathbf{E}}_\rho } \big)$. And the achieved optimal value of problem $\mathcal{P}2.3$, $\Gamma_p^{min*}$, is the largest generalized eigenvalue of the same matrix pair. Thus, the feasible condition is $\Gamma_p^{min} \leq \Gamma_p^{min*}$.

Within the feasible region, we then resort to   the SDR technique and Charnes-Cooper transformation \cite{liu2014secrecy} to solve problem $\mathcal{P}2.1$.
   Introducing ${\mathbf{\bar A}} = {\mathbf{a}}{{\mathbf{a}}^H},{\mathbf{\bar B}} = {\mathbf{b}}{{\mathbf{b}}^H}$
  and applying SDR technique, problem $\mathcal{P}2.1$ can be relaxed as ($\mathcal{P}$2.4)
\begin{subequations}
\begin{align}
&\underset{{{\mathbf{\bar A}},{\mathbf{\bar B}}}}{\text{max}}&& \frac{{\operatorname{Tr} \left( {{{{\mathbf{\hat H}}}_s}{\mathbf{\bar B}}} \right)}}
{{\operatorname{Tr} \left( {{{\mathbf{B}}_\rho }{\mathbf{\bar A}}} \right) + \sigma _s^2}}  \\
&\text{s. t.}&&\!\!\!\! \operatorname{Tr}\left( {{{\mathbf{C}}_\rho }{\mathbf{\bar A}}} \right) \!-\! \Gamma _p^{\min }\left( {\operatorname{Tr} \left( {{{\mathbf{D}}_\rho }{\mathbf{\bar A}}} \right) \!+\! \operatorname{Tr} ( {{{{\mathbf{\hat H}}}_p}{\mathbf{\bar B}}} )} \right) \!\geq\! \Gamma _p^{\min }\sigma _p^2,\\
&&&\operatorname{Tr}\left( {{{\mathbf{E}}_\rho }{\mathbf{\bar A}}} \right) + \operatorname{Tr} \left( {{\mathbf{\bar B}}} \right) \leq P_r^{EH},\\
&&& {\mathbf{\bar A}} \succeq {\mathbf{0}},{\mathbf{\bar B}} \succeq   {\mathbf{0}},
\end{align}
\end{subequations}
  which is a linear fractional quasi-convex problem. From Charnes-Cooper transformation, we define  ${\mathbf{ \hat A}} = q{\mathbf{\bar A}}$, ${\mathbf{ \hat B}} = q{\mathbf{\bar B}}(q > 0)$ and rewrite the problem $\mathcal{P}2.4$ as ($\mathcal{P}2.5$)
\begin{subequations}
\begin{align}
&\underset{{{\mathbf{\hat A}},{\mathbf{\hat B}}, q}}{\text{max}}&& \operatorname{Tr} \left( {{{{\mathbf{\hat H}}}_s}{\mathbf{\hat B}}} \right)
  \\
&\text{s. t.}&& \operatorname{Tr} \left( {{{\mathbf{B}}_\rho }{\mathbf{\hat A}}} \right) + \sigma _s^2q = 1,\\
&&&\!\!\!\!\!\!\!\!\!\!\!\!\!\!\!\!\!\!\!\!\!\!\!\operatorname{Tr} \left( {{{\mathbf{C}}_\rho }{\mathbf{\hat A}}} \right) \!-\! \Gamma _p^{\min }\left( {\operatorname{Tr} ( {{{\mathbf{D}}_\rho }{\mathbf{\hat A}}} )\! + \! \operatorname{Tr} ( {{{{\mathbf{\hat H}}}_p}{\mathbf{\hat B}}} )} \right) \!\geq \!\Gamma _p^{\min }\sigma _p^2q,\!\\
&&& \operatorname{Tr} \left( {{{\mathbf{E}}_\rho }{\mathbf{\hat A}}} \right) + \operatorname{Tr} \left( {{\mathbf{\hat B}}} \right) \leq P_r^{EH}q,\\
&&&{\mathbf{\hat A}} \succeq {\mathbf{0}}, {\mathbf{\hat B}} \succeq {\mathbf{0}}, q > 0,
\end{align}
\end{subequations}
which is a convex SDP and can be efficiently solved by  convex optimization solvers, e.g., CVX \cite{grantmay}.

\textit{ \textbf{Remark 1}:
More importantly, According to the Theorem 2.3 in \cite{huang2010rank}, the optimal solution to problem $\mathcal{P}$2.5 always satisfies  ${\operatorname{Rank} ^2}({{\mathbf{\hat A}}^*}) + {\operatorname{Rank} ^2}({{\mathbf{\hat B}}^*}) \leq 3$, since the number of generalized constraints is 3. For the nontrivial case where  ${{\mathbf{\hat A}}^*} \ne {\mathbf{0}}$, ${{\mathbf{\hat B}}^*} \ne {\mathbf{0}}$, we have  $\operatorname{Rank} ({{\mathbf{\hat A}}^*}) = 1$, $\operatorname{Rank} ({{\mathbf{\hat B}}^*}) = 1$. So the SDR problem is tight and thus the optimal $\mathbf{a}^*$ and $\mathbf{b}^*$ for problem $\mathcal{P}2.1$ can be obtained.}

So far, the optimal solution to problem $\mathcal{P}1$ with fixed $\rho$ is derived. In the sequel, we focus on the finding of optimal $\rho$.

\textit{ \textbf{Proposition 2}:
Define the objective value of problem $\mathcal{P}2.5$ as a function of $\rho$, i.e., $h(\rho)$. Then, $h(\rho)$ is concave in $\rho$ and its optimal value can be obtained via bisection. }

\textit{\textbf{Proof}}:
 Let $\theta_1$, $\theta_2$ and $\theta_3$
  denote the dual variables of the corresponding constraints in problem $\mathcal{P}$2.5, respectively. Then the Lagrangian function of problem $\mathcal{P}$2.5 is given by
\begin{equation}
{\mathcal{L}}\left( {{\mathbf{\hat A}},{\mathbf{\hat B}},q,{\theta _1},{\theta _2},{\theta _3},\rho } \right){\text{ }}=\operatorname{Tr} \left( {{{\mathbf{Q}}_1}{\mathbf{\hat A}}} \right) + \operatorname{Tr} \left( {{{\mathbf{Q}}_2}{\mathbf{\hat B}}} \right) + {q_3},
\end{equation}
where
\begin{equation}
{{\mathbf{Q}}_1} = -{\theta _1}{{\mathbf{B}}_\rho } + {\theta _2}{{\mathbf{C}}_\rho } - {\theta _2}\Gamma _p^{\min } {{\mathbf{D}}_\rho } - {\theta _3}{{\mathbf{E}}_\rho },
\end{equation}
\begin{equation}
{{\mathbf{Q}}_2} = {{{\mathbf{\hat H}}}_s} - {\theta _2}\Gamma _p^{\min }{{{\mathbf{\hat H}}}_p} - {\theta _3}{\mathbf{I}},
\end{equation}
\begin{equation}
\begin{split}
{q_3} = {\theta _1}-{\theta _1}\sigma _s^2q  - {\theta _2}\Gamma _p^{\min }\sigma _p^2q + {\theta _3}q P_r^{EH}.
\end{split}
\end{equation}

The Lagrangian dual function is given by
\begin{equation}
g({\theta _1},{\theta _2},{\theta _3},\rho) = \mathop {\max }\limits_{{\mathbf{\hat A}} \succeq \mathbf{0},{\mathbf{\hat B}} \succeq \mathbf{0},q > 0} \mathcal{L}\left({{\mathbf{\hat A}},{\mathbf{\hat B}},q,{\theta _1},{\theta _2},{\theta _3},\rho }\right).
\end{equation}
Since $\mathcal{P}$2.5 is a convex problem and satisfies the slater's condition, the strong duality holds \cite{boyd2004convex}. Thus, $h({\rho}) = \mathop {\min }\limits_{{\theta _1},{\theta _2}\geq 0,{\theta _3}\geq 0 } g({\theta _1},{\theta _2},{\theta _3},\rho)$.

 From (27)-(30) and (3), we can observe that only $\mathbf{B}_{\rho}$, $\mathbf{C}_{\rho}$, $\mathbf{D}_{\rho}$, $\mathbf{E}_{\rho}$ in $\mathbf{Q}_{1}$ and $P_r^{EH}$ in $q_3$ are related with $\rho$. And all of these terms are linear in $\rho$, such that $g({\theta _1},{\theta _2},{\theta _3},\rho)$ is a linear function  with respect to  $\rho$.
 Accordingly, it is easily verified that  $h({\rho})$ is a point-wise minimum of a family of affine function and hence concave in $\rho$ \cite{boyd2004convex}. Therefore,
its maximum can be found through a one-dimensional search, such as bisection. This completes the proposition. \IEEEQED

As analyzed before, we have $h({\rho}) = \mathcal{L}\left( {{\mathbf{\hat A^*}},{\mathbf{\hat B^*}},q^*,{\theta _1^*},{\theta _2^*},{\theta _3^*},\rho } \right)$, where  ${\mathbf{\hat A^*}},{\mathbf{\hat B^*}},q^*$
are the optimal primary variables and ${\theta _1^*},{\theta _2^*},{\theta _3^*}$ are the optimal dual variables for a given $\rho$, respectively. With (27), (17)-(20) and (3), the gradient of $h({\rho})$ can be expressed as
\begin{align}
\frac{{dh({\rho})}}
{{d{\rho}}} = &\operatorname{Tr} \big( ( - {\theta _1}{{\mathbf{B}}_\rho }_1 + {\theta _2}{{\mathbf{C}}_{\rho 1}} - {\theta _2}\Gamma _p^{\min }{{\mathbf{D}}_{\rho 1}} \\
&- {\theta _3}){\mathbf{\hat A}}{{\mathbf{E}}_{\rho 1}} \big) - {\theta _3}qP_{r1}^{EH}, \nonumber
\end{align}
where
\begin{equation}
{{\mathbf{B}}_{\rho 1}} =  \left( {\sigma _r^2{\mathbf{I}} + {P_{PT}}{{{\mathbf{\bar G}}}^T} + {P_{PR}}{\mathbf{\bar H}}_p^T} \right) \otimes {{{\mathbf{\hat H}}}_s},
\end{equation}
\begin{equation}
{{\mathbf{C}}_{\rho 1}} = {P_{PT}}{{{\mathbf{\bar G}}}^T} \otimes {{{\mathbf{\hat H}}}_p},
\end{equation}
\begin{equation}
{{\mathbf{D}}_{\rho 1}} = \left( {\sigma _r^2{\mathbf{I}} + {P_{SR}}{\mathbf{\bar H}}_s^T} \right) \otimes {{{\mathbf{\hat H}}}_p},
\end{equation}
\begin{equation}
{{\mathbf{E}}_{\rho 1}} = \left( {{P_{PT}}{{{\mathbf{\bar G}}}^T} + {P_{PR}}{\mathbf{\bar H}}_p^T + {P_{SR}}{\mathbf{\bar H}}_s^T} \right) \otimes {{\mathbf{I}}_{2 \times 2}} + \sigma _r^2{{\mathbf{I}}_{6 \times 6}},
\end{equation}
\begin{equation}
P_{r1}^{EH} = \xi ( {{P_{PT}}\left\| {\mathbf{g}} \right\|_2^2 + {P_{{P} R}}\left\| {{{\mathbf{h}}_p}} \right\|_2^2 + {P_{SR}}\left\| {{{\mathbf{h}}_s}} \right\|_2^2 + \sigma _r^2} ).
\end{equation}

Above all, problem $\mathcal{P}$1 can be solved in two steps: (i) Given any
$0<\rho \leq 1$, we first solve the Problem $\mathcal{P}$2.5 to obtain $h({\rho})$; (ii) Then, we use the bisection method to find optimal $\rho$ by using the gradient of $h(\rho)$. Repeat these  two procedures until problem converges. Detailed steps of proposed algorithm are outlined in Algorithm 1. It is worth pointing out that the global optimization  solution to the problem $\mathcal{P}$1 can be achieved by Algorithm 1.

\begin{algorithm}[h]
    \caption{ Optimal solution to problem $\mathcal{P}$1 with perfect CSI}
    \begin{algorithmic}[1]
    \STATE {Initialize ${\rho}^{\min}$, ${\rho}^{\max}$ and tolerance $\delta_{\rho}$;}
     \WHILE {$ {{\rho}^{\max}-{\rho}^{\min} > \delta_t } $}
    \STATE ${\rho} \leftarrow ({\rho}^{\min}+{\rho}^{\max})/2$;
    \STATE Solve problem $\mathcal{P}$2.5  via CVX to obtain ${\mathbf{\hat A^*}},{\mathbf{\hat B^*}},q^*,{\theta _1^*},{\theta _2^*}$ and ${\theta _3^*}$;
    \STATE Calculate $\frac{{dh({\rho})}}{{d{\rho}}}$ according to (32);
    \IF{$\frac{{dh({\rho})}}{{d{\rho}}} \geq 0$}
    \STATE ${\rho}^{\min} \leftarrow {\rho} $;
    \ELSE
    \STATE ${\rho}^{\max} \leftarrow {\rho} $;
    \ENDIF
    \ENDWHILE
    \RETURN  ${\mathbf{\bar A^*}} = {\mathbf{\hat A^*}}/{q^*}$, $\mathbf{\bar B^*} = {\mathbf{\hat B^*}}/ {q^*}$;
    \RETURN $\mathbf{a^*}$ and $\mathbf{b^*}$  via eigenvalue decomposition (EVD) of ${\mathbf{\bar A^*}}$ and ${\mathbf{\bar B^*}}$, respectively;
    \label{code:recentEnd}
    \end{algorithmic}
    \end{algorithm}

\subsection{Suboptimal Solution to DWPT}
Although the optimal solution to problem $\mathcal{P}1$ is obtained, the complexity is high due to the adoption of standard tool box, CVX. In this subsection, we present a low-complexity suboptimal solution, where the closed-form of $\mathbf{F}$ and  $\mathbf{w}$ is derived with given $\rho$. Similar to the former subsection, optimal $\rho$ is found via bisection. In the following, we put the emphasis on the design of $\mathbf{F}$ and  $\mathbf{w}$.

For simplicity, we decompose $\mathbf{F}$ as ${\mathbf{F}} = {{\mathbf{f}}_t}{\mathbf{f}}_r^H$ \cite{zheng2015joint}, where $\mathbf{f}_t$ is the transmit beamforming vector and $\mathbf{f}_r$ is receiver filter at ST. Without loss of generality, we further suppose that $\left\| {{{\mathbf{f}}_r}} \right\|_2^2 = 1$. According to the propertity of  matrix norm
\begin{equation}
\left\| {{{\mathbf{f}}_t}{\mathbf{f}}_r^H} \right\|_F^2 \leq \left\| {{{\mathbf{f}}_t}} \right\|_2^2\left\| {{\mathbf{f}}_r^H} \right\|_2^2,
 \end{equation}
 the original problem $\mathcal{P}1$ is converted as (with fixed $\rho$) ($\mathcal{P}3$)
\begin{subequations}
\begin{align}
&\!\underset{{{\mathbf{f}_t},{\mathbf{w}},\atop \left\| {{{\mathbf{f}}_r}} \right\|_2^2 = 1}}{\text{max}}&&\!\!\! \frac{{{{\left| {{\mathbf{h}}_s^H{\mathbf{w}}} \right|}^2}}}
{{{{\left| {{\mathbf{h}}_s^H{{\mathbf{f}}_t}} \right|}^2}(\rho ({P_{PT}}{{\left| {{\mathbf{f}}_r^H{\mathbf{g}}} \right|}^2} \!\!+ \!\!{P_{PR}}{{\left| {{\mathbf{f}}_r^H{{\mathbf{h}}_p}} \right|}^2}) \!\!+\!\! (\rho \sigma _r^2\!\! +\!\! \sigma _c^2)) \!\!+\!\! \sigma _s^2}} \displaybreak[0] \\
&\text{s. t.}&&\!\!\!\!\!\!\!\!\!\!  \frac{{\rho {P_{PT}}{{\left| {{\mathbf{h}}_p^H{{\mathbf{f}}_t}} \right|}^2}{{\left| {{\mathbf{f}}_r^H{\mathbf{g}}} \right|}^2}}}{{{{\left| {{\mathbf{h}}_p^H{{\mathbf{f}}_t}} \right|}^2}(\rho {P_{SR}}{{\left| {{\mathbf{f}}_r^H{\mathbf{h}}_s^{}} \right|}^2}\!\! +\!\! (\rho \sigma _r^2\!\! +\!\! \sigma _c^2))\!\! +\!\! {{\left| {{\mathbf{h}}_p^H{\mathbf{w}}} \right|}^2} \!\!+ \!\! \sigma _p^2}}\!\! \geq \!\!\Gamma _p^{\min }, \displaybreak[0]\\
 &&&\!\!\!\!\!\!\!\!\!\!\left\| {{{\mathbf{f}}_t}} \right\|_2^2\big(\rho ({P_{PT}}{\left| {{\mathbf{f}}_r^H{\mathbf{g}}} \right|^2} + {P_{PR}}{\left| {{\mathbf{f}}_r^H{{\mathbf{h}}_p}} \right|^2} + {P_{SR}}{\left| {{\mathbf{f}}_r^H{{\mathbf{h}}_s}} \right|^2}) \nonumber  \displaybreak[0]\\
 &&&+ (\rho \sigma _r^2 + \sigma _c^2)\big) + \left\| {\mathbf{w}} \right\|_2^2  \leq P_r^{EH}, \displaybreak[0]
\end{align}
\end{subequations}
which is non-convex due to coupling variables $\mathbf{f}_t$ and $\mathbf{f}_r$.

To tackle this difficulty effectively, we first design $\mathbf{f}_r$ with given $\mathbf{f}_t$ and $\mathbf{w}$. The ZF method is used here. Suppose that ${\mathbf{h}}_s^H{{\mathbf{f}}_r}\!\! =\!\! 0$ and ${\mathbf{h}}_p^H{{\mathbf{f}}_r} \!\!=\!\! 0$, we have  ${{\mathbf{f}}_r} \!\!=\!\! {\mathbf{\tilde V}}{{\mathbf{\tilde f}}_r}$, where ${\mathbf{\tilde V}}$ is the orthogonal basis for null space of  $\left[ {{{\mathbf{h}}_s},{{\mathbf{h}}_p}} \right]^H$
 and can be derived from the singular value decomposition (SVD) method \cite{liu2014secrecy}. Then, to satisfy the SINR of PR, we further assume that $\tilde {\mathbf{f}}_r$ is aligned to the same direction of  ${{\mathbf{\tilde V}}^H}{\mathbf{g}}$. Together with  $\left\| {{{\mathbf{f}}_r}} \right\|_2^2 = 1$, we have
 \begin{equation}
 {{\mathbf{f}}_r} \!\!=\!\! {\mathbf{\tilde V}}\frac{{{{{\mathbf{\tilde V}}}^H}{\mathbf{g}}}}
{{\left\| {{{{\mathbf{\tilde V}}}^H}{\mathbf{g}}} \right\|_2^2}}.
\end{equation}

With fixed $\mathbf{f}_r$, the problem $\mathcal{P}3$ is briefly expressed as ($\mathcal{P}3.1$)
\begin{subequations}
\begin{align}
&\underset{{{\mathbf{f}_t},{\mathbf{w}}}}{\text{max}}&& \frac{{{{\left| {{\mathbf{h}}_s^H{\mathbf{w}}} \right|}^2}}}
{{b{{\left| {{\mathbf{h}}_s^H{{\mathbf{f}}_t}} \right|}^2} + \sigma _s^2}} \\
&\text{s. t.}&&\!\!\!\!\!\!\!\!\! \!\! \frac{{{{\left| {{\mathbf{h}}_p^H{{\mathbf{f}}_t}} \right|}^2}}}
{{c{{\left| {{\mathbf{h}}_p^H{{\mathbf{f}}_t}} \right|}^2} + {{\left| {{\mathbf{h}}_p^H{\mathbf{w}}} \right|}^2} + \sigma _p^2}} \geq \Gamma _{pt}, \displaybreak[0]\\
&&& d\left\| {{{\mathbf{f}}_t}} \right\|_2^2 + \left\| {\mathbf{w}} \right\|_2^2 \leq P_r^{EH},
\end{align}
\end{subequations}
where
\begin{equation}
{\text{b = }}\rho {P_{PT}}{\left| {{\mathbf{f}}_r^H{\mathbf{g}}} \right|^2} + \rho {P_{PR}}{\left| {{\mathbf{f}}_r^H{{\mathbf{h}}_p}} \right|^2} + (\rho \sigma _r^2 + \sigma _c^2),
 \end{equation}
 \begin{equation}
 c = \rho {P_{SR}}{\left| {{\mathbf{f}}_r^H{\mathbf{h}}_s^{}} \right|^2} + (\rho \sigma _r^2 + \sigma _c^2),
 \end{equation}
 \begin{equation}
 {\Gamma _{pt}} = \frac{\Gamma _p^{\min }}  {\rho {P_{PT}}{{\left| {{\mathbf{f}}_r^H{\mathbf{g}}} \right|}^2}},~\text{and}
 \end{equation}
  \begin{equation}
  d \!= \!\rho({P_{PT}} {\left| {{\mathbf{f}}_r^H{\mathbf{g}}} \right|^2} \!+\! {P_{PR}} {\left| {{\mathbf{f}}_r^H{{\mathbf{h}}_p}} \right|^2}\! + \!{P_{SR}}{\left| {{\mathbf{f}}_r^H{{\mathbf{h}}_s}} \right|^2})\! + \!(\rho \sigma _r^2 \!+ \!\sigma _c^2).
  \end{equation}
It can be easily verified that, at the optimum,  constraints  (41b) and (41c) are all active. In particular, if the power constraint (41c) is not active at the optimum, we can increase the power of cognitive beamforming vector $\mathbf{w}$ in the null space of $\mathbf{h}_p$ until that (41c) is active. In this way, the objective value is increasing while the constraint (41b) remains unchanged, which contradicts to the optimality point assumption. Then, if the constraint (41b) is not active, we can keep the direction of the transmit vector $\mathbf{f}_t$ unchanged and  decrease its transmission power such that (41b) is active. During this process, the constraint (41c) becomes non-active  and the objective value of problem $\mathcal{P}3.1$ is increasing, which also contradicts to the optimality point assumption. Thus, at the optimum, both (41b) and (41c) are active.

 To obtain the closed-form solution, we first consider the power minimization problem with satisfied rates of PU and SU as follows ($\mathcal{P}3.2$):
\begin{subequations}
\begin{align}
&\underset{{{\mathbf{f}_t},{\mathbf{w}}}}{\text{min}}&& d\left\| {{{\mathbf{f}}_t}} \right\|_2^2 + \left\| {\mathbf{w}} \right\|_2^2 \\
&\text{s. t.}&&\!\!\!\!\!  \frac{{{{\left| {{\mathbf{h}}_p^H{{\mathbf{f}}_t}} \right|}^2}}}
{c{{\left| {{\mathbf{h}}_p^H{{\mathbf{f}}_t}} \right|}^2} + {{\left| {{\mathbf{h}}_p^H{\mathbf{w}}} \right|}^2 + \sigma _p^2}} \geq {\Gamma _{pt}}, \\
&&&~~~\frac{{{{\left| {{\mathbf{h}}_s^H{\mathbf{w}}} \right|}^2}}}
{{b{{\left| {{\mathbf{h}}_s^H{{\mathbf{f}}_t}} \right|}^2} + \sigma _s^2}} \geq {\Gamma _{st}},
\end{align}
\end{subequations}
where $\Gamma_{st}$ is viewed as the optimal value of problem $\mathcal{P}3.1$.
In the similar way to the proof of problem $\mathcal{P}3.1$, one can also prove that, constraints  (46b) and (46c) are all active at the optimum. Thus, we are sure that the optimal objective value of problem $\mathcal{P}3.2$ is exactly $P_r^{EH}$. 

According to \cite{boyd2004convex}, the dual problem of $\mathcal{P}3.2$ is expressed as ($\mathcal{P}3.3$)
\begin{subequations}
\begin{align}
&\underset{{{\lambda_1\geq 0},{\lambda_2 \geq 0}}}{\text{max}}&& ~~\sigma _p^2{\lambda _1} + \sigma _s^2{\lambda _2} \\
&\text{s. t.}&&\!\!\!\!\! {\mathbf{I}} + {\lambda _1}{\mathbf{h}}_p^{}{\mathbf{h}}_p^H \succeq   \frac{{{\lambda _2}}}
{{{\Gamma _{st}}}}{\mathbf{h}}_s^{}{\mathbf{h}}_s^H, \\
 &&&\!\!\!\!\!{\mathbf{I}} + \frac{{b{\lambda _2}}}
{d}{\mathbf{h}}_s^{}{\mathbf{h}}_s^H \succeq  e{\lambda _1}{\mathbf{h}}_p^{}{\mathbf{h}}_p^H,
\end{align}
\end{subequations}
where $\lambda_1$, $\lambda_2$ are dual variables and $e = \frac{{1 - c{\Gamma _{pt}}}}
{{d{\Gamma _{pt}}}}$. From the  constraint  (41b), we know that, if  ${\left| {{\mathbf{h}}_p^H{{\mathbf{f}}_t}} \right|^2} \to \infty$, $\frac{{{{\left| {{\mathbf{h}}_p^H{{\mathbf{f}}_t}} \right|}^2}}}
{{c{{\left| {{\mathbf{h}}_p^H{{\mathbf{f}}_t}} \right|}^2} + {{\left| {{\mathbf{h}}_p^H{\mathbf{w}}} \right|}^2} + \sigma _p^2}} \to \frac{1}
{c}$. Since the non-trivial case, where the problem $\mathcal{P}3.1$ is feasible, is considered, we have
\begin{equation}
{\Gamma _{pt}} \!\leq\! {\left. {\frac{{{{\left| {{\mathbf{h}}_p^H{{\mathbf{f}}_t}} \right|}^2}}}
{{c{{\left| {{\mathbf{h}}_p^H{{\mathbf{f}}_t}} \right|}^2}\! + \!{{\left| {{\mathbf{h}}_p^H{\mathbf{w}}} \right|}^2} \!+\! \sigma _p^2}}} \right|_{{{\mathbf{f}}_t} = \frac{{P_r^{EH}}}
{d}{\mathbf{h}}_p^{},{\mathbf{w}} = {\mathbf{0}}}}\! < \!\frac{1}
{c}.
\end{equation} Thus, $c{\Gamma _{pt}} < 1$, i.e., $e>0$ must hold. From (42) and (45), we also have $b>0$ and $d>0$.

Since ${\mathbf{I}} + {\lambda _1}{\mathbf{h}}_p^{}{\mathbf{h}}_p^H \succ
\mathbf{0}$, constraint (47b) can be rewritten as
\begin{equation}
{\mathbf{I}} \succeq   \frac{{{\lambda _2}}}
{{{\Gamma _{st}}}}{\left( {{\mathbf{I}} + {\lambda _1}{\mathbf{h}}_p^{}{\mathbf{h}}_p^H} \right)^{ - 1}}{\mathbf{h}}_s^{}{\mathbf{h}}_s^H.
\end{equation}
Due to the fact that  $\operatorname{Rank} \left( {{{({\mathbf{I}} + {\lambda _1}{\mathbf{h}}_p^{}{\mathbf{h}}_p^H)}^{ - 1}}{\mathbf{h}}_s^{}{\mathbf{h}}_s^H} \right) = 1$, this matrix only has one nonnegative eigenvalue, that is, ${\mathbf{h}}_s^H{\left( {{\mathbf{I}} + {\lambda _1}{\mathbf{h}}_p^{}{\mathbf{h}}_p^H} \right)^{ - 1}}{\mathbf{h}}_s^{}$. Hence, (49) is equivalent to
\begin{equation}
{\Gamma _{st}} \geq {\lambda _2}{\mathbf{h}}_s^H{\left( {{\mathbf{I}} + {\lambda _1}{\mathbf{h}}_p^{}{\mathbf{h}}_p^H} \right)^{ - 1}}{\mathbf{h}}_s^{}.
\end{equation}
Similarly, (47c) can be equivalently reformulated as
\begin{equation}
1 \geq e{\lambda _1}{\mathbf{h}}_p^H{( {{\mathbf{I}} + \frac{{b{\lambda _2}}}
{d}{\mathbf{h}}_s^{}{\mathbf{h}}_s^H} )^{ - 1}}{\mathbf{h}}_p^{}.
\end{equation}
To maximize the objective of problem $\mathcal{P}3.3$, i.e., $\sigma _p^2{\lambda _1} + \sigma _s^2{\lambda _2}$, one can easily verify that, at the optimum,  two constraints (50) and (51) are both active, i.e.,
\begin{equation}
{\Gamma _{st}} = {\lambda _2}{\mathbf{h}}_s^H{\left( {{\mathbf{I}} + {\lambda _1}{\mathbf{h}}_p^{}{\mathbf{h}}_p^H} \right)^{ - 1}}{\mathbf{h}}_s^{}~\text{and}
 \end{equation}
 \begin{equation}
 1 = e{\lambda _1}{\mathbf{h}}_p^H{\left( {\mathbf{I}} + {\frac{b{\lambda _2}} {d}{\mathbf{h}}_s^{}{\mathbf{h}}_s^H} \right)^{ - 1}}{\mathbf{h}}_p^{}.
   \end{equation}
 Then,  using the matrix inversion lemma
 \begin{equation}
 {\left( {{\mathbf{A}} + {\mathbf{u}}{{\mathbf{v}}^H}} \right)^{ - 1}} = {{\mathbf{A}}^{ - 1}} - \frac{{{{\mathbf{A}}^{ - 1}}{\mathbf{u}}{{\mathbf{v}}^H}{{\mathbf{A}}^{ - 1}}}}
{{1 + {{\mathbf{v}}^H}{{\mathbf{A}}^{ - 1}}{\mathbf{u}}}},
\end{equation}
  we have
\begin{eqnarray}
&&\!\!\!\!\!\!\!\!{\lambda _2} = \frac{{{\Gamma _{st}}\left( {1 + {\lambda _1}{{\left\| {{\mathbf{h}}_p^{}} \right\|}_2^2}} \right)}}
{{{{\left\| {{\mathbf{h}}_s^{}} \right\|}_2^2} + {\lambda _1}\left( {{{\left\| {{\mathbf{h}}_s^{}} \right\|}_2^2}{{\left\| {{\mathbf{h}}_p^{}} \right\|}_2^2} - {{\left| {{\mathbf{h}}_s^H{\mathbf{h}}_p^{}} \right|}^2}} \right)}}, \\
&&\!\!\!\!\!\!\!\! {\lambda _1} = \frac{{\frac{1}
{e} + \frac{{b{\lambda _2}}}
{{ed}}{{\left\| {{\mathbf{h}}_s^{}} \right\|}^2_2}}}
{{{{\left\| {{\mathbf{h}}_p^{}} \right\|}^2_2} + \frac{{b{\lambda _2}}}
{d}\left( {{{\left\| {{\mathbf{h}}_s^{}} \right\|}^2_2}{{\left\| {{\mathbf{h}}_p^{}} \right\|}^2_2} - {{\left| {{\mathbf{h}}_s^H{\mathbf{h}}_p^{}} \right|}^2}} \right)}}.
\end{eqnarray}

As mentioned before, the optimal objective value of problem $\mathcal{P}3.2$ is $P_r^{EH}$. Due to the strong duality,  the objective value of its dual problem must satisfy
\begin{equation}
\sigma _p^2{\lambda _1} + \sigma _s^2{\lambda _2} = P_r^{EH}.
 \end{equation}
 Combining this power equation (57) with (56), a quadratic function in terms of $\lambda_2$ is derived. That is,
 \begin{equation}
 f({\lambda _2}) = \alpha \lambda _2^2 + \beta {\lambda _2} + \gamma  = 0,
 \end{equation} where
 \begin{equation}
 \alpha  = \frac{b}
{d}\sigma _s^2\left( {{{\left\| {{\mathbf{h}}_s^{}} \right\|}^2_2}{{\left\| {{\mathbf{h}}_p^{}} \right\|}^2_2} - {{\left| {{\mathbf{h}}_s^H{\mathbf{h}}_p^{}} \right|}^2}} \right) > 0,
 \end{equation}
 \begin{equation}
 \beta \! =\! \sigma _s^2{\left\| {{\mathbf{h}}_p^{}} \right\|^2_2} - P_r^{EH}\frac{b}{d}\left( {{{\left\| {{\mathbf{h}}_s^{}} \right\|}^2_2}{{\left\| {{\mathbf{h}}_p^{}} \right\|}^2_2}\! -\! {{\left| {{\mathbf{h}}_s^H{\mathbf{h}}_p^{}} \right|}^2}} \right) + \frac{{b\sigma _p^2}}{{ed}}{\left\| {{\mathbf{h}}_s^{}} \right\|^2_2},
 \end{equation}
 \begin{equation}
 \text{and}~\gamma  = \frac{{\sigma _p^2}}
{e} - P_r^{EH}{\left\| {{\mathbf{h}}_p^{}} \right\|^2_2} < 0.
 \end{equation}
 $\gamma<0$ is obtained from  (48). Note that  $\gamma=0$ is the trivial case where the SU rate is zero.
Therefore, $f({\lambda _2}) = 0$ always has one unique real root, which can be derived based on the roots formula of the quadric equation. $\lambda_1^*$ and $\Gamma_{st}^*$ can then be obtained based on (55) and (56). Thus, the optimal value of problem $\mathcal{P}3.1$ is achieved.

Meanwhile, the closed-form of $\mathbf{f}_t^*$ and $\mathbf{w}^*$ is given by
\begin{equation}
 {{\mathbf{f}}_t^*} = \sqrt {p_{{f_t}}^*} {{\mathbf{\hat f}}_t} = \sqrt {p_{{f_t}}^*} \frac{{{{\left( {{\mathbf{I}} + \frac{{b{\lambda _2}}}
{d}{\mathbf{h}}_s^{}{\mathbf{h}}_s^H} \right)}^{ - 1}}{\mathbf{h}}_p^{}}}
{{\left\| {{{\left( {{\mathbf{I}} + \frac{{b{\lambda _2}}}
{d}{\mathbf{h}}_s^{}{\mathbf{h}}_s^H} \right)}^{ - 1}}{\mathbf{h}}_p^{}} \right\|_2^2}},
\end{equation}
\begin{equation}
{{\mathbf{w}}^*} = \sqrt {p_w^*} {\mathbf{\hat w}} = \sqrt {p_w^*} \frac{{{{\left( {{\mathbf{I}} + {\lambda _1}{\mathbf{h}}_p^{}{\mathbf{h}}_p^H} \right)}^{ - 1}}{\mathbf{h}}_s^{}}}
{{\left\| {{{\left( {{\mathbf{I}} + {\lambda _1}{\mathbf{h}}_p^{}{\mathbf{h}}_p^H} \right)}^{ - 1}}{\mathbf{h}}_s^{}} \right\|_2^2}},
\end{equation}
where the power $p_{{f_t}}^*$, $p_w^*$ can be easily obtained from
\begin{equation}
\left\{ \begin{aligned}
  &\frac{{p_{{f_t}}^*{{\left| {{\mathbf{h}}_p^H{{{\mathbf{\hat f}}}_t}} \right|}^2}}}{{cp_{{f_t}}^*{{\left| {{\mathbf{h}}_p^H{{{\mathbf{\hat f}}}_t}} \right|}^2} + p_w^*{{\left| {{\mathbf{h}}_p^H{\mathbf{\hat w}}} \right|}^2} + \sigma _p^2}} = {\Gamma _{pt}}, \\
 &dp_{{f_t}}^* + p_w^* = P_r^{EH}.
 \end{aligned} \right.
\end{equation}

To summarize, the proposed  low-complexity suboptimal solution to problem $\mathcal{P}1$ consists of two steps: (i) With a given $0< \rho \leq 1$, $\mathbf{f}_r$ can be first derived based on (40). Then, $\mathbf{f}_t$ and $\mathbf{w}$ are obtained based on (62) and (63), respectively. Note that the expression of  $\mathbf{f}_r$ is  independent of  $\mathbf{f}_t$ and $\mathbf{w}$, so it is not necessary to optimize $\mathbf{f}_r$ and   ($\mathbf{f}_t$, $\mathbf{w}$) iteratively. (ii) Similar to the optimal solution in subsection A,  the optimal $\rho$ can be found via bisection. Repeat these two procedures until problem converges.

\section{Robust Solution to DWPT with Imperfect CSI}

Until now, we assume that all channel state information is perfectly known at ST. In practice, the knowledge of channel $\mathbf{g}$ can be directly estimated by ST, whereas the CSI of $\mathbf{h}_p$/$\mathbf{h}_s$ depends on the quantized feedback from PR/SR. As a result, the level of channel uncertainty is much higher in $\mathbf{h}_p$ and $\mathbf{h}_s$. For this reason, the perfect CSI of $\mathbf{g}$ and imperfect CSI of $\mathbf{h}_p$ and $\mathbf{h}_s$ are considered in this section \cite{ponukumati2013robust}. 

The imperfect $\mathbf{h}_p$ and $\mathbf{h}_s$ are respectively modeled as
\begin{equation}
{{\mathbf{h}}_p} = {{\mathbf{\tilde h}}_p} + \Delta {{\mathbf{h}}_p},{\mathcal{B}_p} = \left\{ {\Delta {{\mathbf{h}}_p}:\left\| {\Delta {{\mathbf{h}}_p}} \right\|_2^2 \leq \varepsilon _p^2} \right\}~~ \text{and}
\end{equation}
\begin{equation}
{{\mathbf{h}}_s} = {{\mathbf{\tilde h}}_s} + \Delta {{\mathbf{h}}_s},{\mathcal{B}_s} = \left\{ {\Delta {{\mathbf{h}}_s}:\left\| {\Delta {{\mathbf{h}}_s}} \right\|_2^2 \leq \varepsilon _s^2} \right\}
\end{equation}
where $\tilde {\mathbf{h}}_p$ and $\tilde {\mathbf{h}}_s$ are the estimated CSI; $\Delta {{\mathbf{h}}_p}$ and $\Delta {{\mathbf{h}}_s}$ are the channel error vectors; ${\varepsilon _p^2}$ and ${\varepsilon _s^2}$ are radii of the channel error uncertainty regions.

With the imperfect CSI, the harvested energy at ST (i.e., (3)) and the transmission power at ST (i.e., (6)) can be respectively rewritten as
\begin{align}
\tilde P_r^{EH} =& \xi (1 - \rho )\bigg( {P_{PR}}\left\| {{{{\mathbf{\tilde h}}}_p} + \Delta {{\mathbf{h}}_p}} \right\|_2^2 + {P_{SR}}\left\| {{{{\mathbf{\tilde h}}}_s} + \Delta {{\mathbf{h}}_s}} \right\|_2^2 \nonumber \\
&+ {P_{PT}}\left\| {\mathbf{g}} \right\|_2^2 + \sigma _r^2 \bigg)~ \text{and}
\end{align}
\begin{align}
  &\!\!\!\!{{\tilde P}_{ST}}({\mathbf{F}},{\mathbf{w}},\rho )\! =\! \rho \bigg(\!\!{P_{PT}}\left\| {{\mathbf{Fg}}} \right\|_2^2 + {P_{PR}} \left\| {{\mathbf{F}}({{{\mathbf{\tilde h}}}_p} + \Delta {{\mathbf{h}}_p})} \right\|_2^2 \\
  &\!\!\!\!+ {P_{SR}} \left\| {{\mathbf{F}}({{{\mathbf{\tilde h}}}_s} + \Delta {{\mathbf{h}}_s})} \right\|_2^2 + \sigma _r^2\left\| {\mathbf{F}} \right\|_F^2\bigg) + \sigma _c^2\left\| {\mathbf{F}} \right\|_F^2 + \left\| {\mathbf{w}} \right\|_2^2. \nonumber
\end{align}

It is worth pointing out that, with the imperfect $\mathbf{h}_p$ and $\mathbf{h}_s$, the self-interference of the received signal at PR/SR cannot be cancelled completely. Actually, only the estimated part of self-interference can be removed. Consequently, the received signals at PR and SR are respectively given as
\begin{align}
{{\tilde y}_p} =& \underbrace {\sqrt \rho  {{( {{\mathbf{\tilde h}}_p^{} + \Delta {\mathbf{h}}_p^{}} )}^H}{\mathbf{Fg}}{x_p}}_{{\text{Desired signal}}}\displaybreak[0]\\
  &+ \underbrace {\sqrt \rho  \left( {\Delta {\mathbf{h}}_p^H{\mathbf{F}}{{({\mathbf{\tilde h}}_p
  + \Delta {\mathbf{h}}_p^{})}} + {\mathbf{\tilde h}}_p^H{\mathbf{F}}\Delta {\mathbf{h}}_p} \right){{x}_p'}}_{{\text{Residual self-interference}}} \nonumber \displaybreak[0]\\
  & + \underbrace { {{( {{\mathbf{\tilde h}}_p^{} + \Delta {\mathbf{h}}_p^{}} )}^H}\left(\sqrt \rho {{\mathbf{F}}( {{\mathbf{\tilde h}}_s^{} + \Delta {\mathbf{h}}_s^{}} ){{x}_s'} + {\mathbf{w}}{x_s}} \right)}_{{\text{Interference caused by SU}}} \nonumber \displaybreak[0]\\
   &+ \underbrace {{{( {{\mathbf{\tilde h}}_p^{} + \Delta {\mathbf{h}}_p^{}} )}^H}{\mathbf{F}}\left( {\sqrt \rho  {{\mathbf{n}}_r} + {{\mathbf{n}}_c}} \right) + {n_p}}_{\text{Noise}} \nonumber  ~\text{and}
\end{align}
\begin{align}
{\tilde y_s} =& \underbrace {{{({\mathbf{\tilde h}}_s^{} + \Delta {\mathbf{h}}_s^{})}^H}{\mathbf{w}}x_s}_{{\text{Desired signal}}} \displaybreak[0] \\
  &+ \underbrace {\sqrt \rho  \left( {\Delta {\mathbf{h}}_s^H{\mathbf{F}}{{({\mathbf{\tilde h}}_s^{} + \Delta {\mathbf{h}}_s^{})}} + {\mathbf{\tilde h}}_s^H{\mathbf{F}}\Delta {\mathbf{h}}_s} \right){{x}_s'}}_{{\text{Residual self-interference}}} \nonumber \displaybreak[0]\\
  &+ \underbrace {\sqrt \rho  {{({\mathbf{\tilde h}}_s^{} + \Delta {\mathbf{h}}_s^{})}^H}{\mathbf{F}}\left( {({\mathbf{\tilde h}}_p^{} + \Delta {\mathbf{h}}_p^{}){{x}_p'} + {\mathbf{g}}{x_p}} \right)}_{{\text{Interference caused by PU}}} \nonumber \displaybreak[0] \\
  &+ \underbrace {{{({\mathbf{\tilde h}}_s^{} + \Delta {\mathbf{h}}_s^{})}^H}{\mathbf{F}}(\sqrt \rho  {{\mathbf{n}}_r} + {{\mathbf{n}}_c}) + {n_s}}_{{\text{Noise}}}. \nonumber
\end{align}

Hence, from (69), the SINR at PR is
\begin{equation}
{{\tilde \Gamma }_p} = \frac{{{\mu _1}}}
{{{\mu _2} + {\mu _3} + {\mu _4}}},
\end{equation}
where
\begin{equation}
{\mu _1} = \rho {P_{PT}}{\left| {{{( {{\mathbf{\tilde h}}_p^{} + \Delta {\mathbf{h}}_p^{}} )}^H}{\mathbf{Fg}}} \right|^2},
\end{equation}
\begin{equation}
{\mu _2} = \rho {P_{{P}R}}{\left| {\Delta {\mathbf{h}}_p^H{\mathbf{F}}({\mathbf{\tilde h}}_p^{} + \Delta {\mathbf{h}}_p^{}) + {\mathbf{\tilde h}}_p^H{\mathbf{F}}\Delta {\mathbf{h}}_p^{}} \right|^2},
\end{equation}
\begin{align}
{\mu _3} = &\rho {P_{SR}}{\left| {{{( {{\mathbf{\tilde h}}_p^{} + \Delta {\mathbf{h}}_p^{}} )}^H}{\mathbf{F}}( {{\mathbf{\tilde h}}_s^{} + \Delta {\mathbf{h}}_s^{}} )} \right|^2}\\
 &+  {\left| {{{( {{\mathbf{\tilde h}}_p^{} + \Delta {\mathbf{h}}_p^{}} )}^H}{\mathbf{w}}} \right|^2}~\text{and} \nonumber
\end{align}
\begin{equation}
{\mu _4} = (\rho \sigma _r^2 + \sigma _c^2)\left\| {{{( {{\mathbf{\tilde h}}_p^{} + \Delta {\mathbf{h}}_p^{}} )}^H}{\mathbf{F}}} \right\|_2^2 + \sigma _p^2.
\end{equation}
Similarly, from (70), the SINR at SR is
\begin{equation}
{{\tilde \Gamma }_s} = \frac{{{\eta _1}}}
{{{\eta _2} + {\eta _3} + {\eta _4}}}
\end{equation}
where
\begin{equation}
{\eta _1} = {\left| {{{( {{\mathbf{\tilde h}}_s^{} + \Delta {\mathbf{h}}_s^{}} )}^H}{\mathbf{w}}} \right|^2},
\end{equation}
\begin{equation}
{\eta _2} = \rho {P_{SR}}{\left| {\Delta {\mathbf{h}}_s^H{\mathbf{F}}{{({\mathbf{\tilde h}}_s^{} + \Delta {\mathbf{h}}_s^{})}^{}} + {\mathbf{\tilde h}}_s^H{\mathbf{F}}\Delta {\mathbf{h}}_s^{}} \right|^2},
\end{equation}
\begin{align}
{\eta _3} = &\rho {P_{PR}}{\left| {{{( {{\mathbf{\tilde h}}_s^{} + \Delta {\mathbf{h}}_s^{}} )}^H}{\mathbf{F}}( {{\mathbf{\tilde h}}_p^{} + \Delta {\mathbf{h}}_p^{}} )} \right|^2} \\
&+ \rho {P_{PT}}{\left| {{{( {{\mathbf{\tilde h}}_s^{} + \Delta {\mathbf{h}}_s^{}} )}^H}{\mathbf{Fg}}} \right|^2} ~\text{and} \nonumber
\end{align}
\begin{equation}
{\eta _4} = (\rho \sigma _r^2 + \sigma _c^2)\left\| {{{( {{\mathbf{\tilde h}}_s^{} + \Delta {\mathbf{h}}_s^{}} )}^H}{\mathbf{F}}} \right\|_2^2 + \sigma _s^2.
\end{equation}

Accordingly, the worst-case SU rate maximization problem subject to PU rate constraint and energy causality constraint  is formulated as ($\mathcal{P}4$)
\begin{subequations}
\begin{align}
&\underset{{{\mathbf{F}},{\mathbf{w}},0\leq\rho\leq 1}}{\text{max}}&& \underset{{\Delta {{\mathbf{h}}_p} \in {\mathcal{B}_p},\Delta {{\mathbf{h}}_s} \in {\mathcal{B}_s}}}{\text{min}}~~\frac{{{\eta _1}}}
{{{\eta _2} + {\eta _3} + {\eta _4}}} \\
&\text{s. t.}&&  \!\!\!\!\!\!\!\!\!\!\!\!\!\!\!\!\!\! \frac{{{\mu _1}}}
{{{\mu _2} + {\mu _3} + {\mu _4}}} \geq \Gamma _p^{\min },\forall \Delta {{\mathbf{h}}_p} \!\in\! {\mathcal{B}_p},\!\forall \Delta {{\mathbf{h}}_s}\! \in\! {\mathcal{B}_s}, \displaybreak[0] \\
&&&\!\!\!\!\!\!\!\!\!\!\!\!\!\!\! \!\!\!\!\!\!\!\!\!\!\!\!\!\!\!\!{{\tilde P}_{ST}}({\mathbf{F}},{\mathbf{w}},\rho ) \leq \tilde P_r^{EH},{\text{ }}\forall \Delta {{\mathbf{h}}_p} \in {\mathcal{B}_p},\forall \Delta {{\mathbf{h}}_s} \in {\mathcal{B}_s}.
\end{align}
\end{subequations}

Referring to \cite{boyd2004convex}, this max-min problem can be equivalently rewritten as ($\mathcal{P}4.1$)
\begin{subequations}
\begin{align}
&\underset{{{\mathbf{F}},{\mathbf{w}},0\leq\rho\leq 1, t}}{\text{max}}&& ~~t\\
&\text{s. t.}&&  \!\!\!\!\!\!\!\!\!\!\!\!\!\!\!\!\!\! {\eta _1} \geq t\left( {{\eta _2} + {\eta _3} + {\eta _4}} \right),\forall \Delta {{\mathbf{h}}_p} \!\in\! {\mathcal{B}_p},\!\forall \Delta {{\mathbf{h}}_s}\! \in\! {\mathcal{B}_s}, \displaybreak[0] \\
&&&\!\!\!\!\!\!\!\!\!\!\!\!\!\!\! \!\!\!\!\!\!\!\!\!\!\!\!\!\!\!\!\!\!\!{\mu _1} \geq \Gamma _p^{\min }({\mu _2} + {\mu _3} + {\mu _4}),{\text{ }}\forall \Delta {{\mathbf{h}}_p} \in {\mathcal{B}_p},\forall \Delta {{\mathbf{h}}_s} \in {\mathcal{B}_s}, \displaybreak[0] \\
&&&\!\!\!\!\!\!\!\!\!\!\!\!\!\!\! \!\!\!\!\!\!\!\!\!\!\!\!\!{{\tilde P}_{ST}}({\mathbf{F}},{\mathbf{w}},\rho ) \leq \tilde P_r^{EH},{\text{ }}\forall \Delta {{\mathbf{h}}_p} \in {\mathcal{B}_p},\forall \Delta {{\mathbf{h}}_s} \in {\mathcal{B}_s},
\end{align}
\end{subequations}
where $t$ is a introduced nonnegative parameter. Given fixed $\rho$ and $t$,  generally speaking, this kind of robust problem  can be solved by the SDR technique and S-procedure \cite{ yu2015optimal,sun2015multi}. Nevertheless, different from the beamforming vectors design for the simple downlink broadcast scenario in \cite{yu2015optimal}, our considered problem involves two hops relay transmission and the design of relay matrix $\mathbf{F}$. To tackle the difficulty caused by $\mathbf{F}$, the matrix properties described in \textbf{Lemma 1} \cite{wurobust} are applied to transform the $\mathbf{F}$ related terms into our desired expressions.

\textit{\textbf{Lemma 1}:Define ${\mathbf{f}} = {\operatorname{vec} }({\mathbf{F}})$, we have
\begin{align}
\Delta {{\mathbf{z}}^T}{\mathbf{Fg}} = \Delta {{\mathbf{z}}^T}({{\mathbf{g}}^T} \otimes {\mathbf{I}}){\mathbf{f}},
\end{align}
\begin{align}
{{\mathbf{g}}^T}{\mathbf{F}}\Delta {\mathbf{z}} = \Delta {{\mathbf{z}}^T}({\mathbf{I}} \otimes {{\mathbf{g}}^T}){\mathbf{f}},
\end{align}
\begin{align}
{{\mathbf{z}}^T}{\mathbf{F}}{{\mathbf{F}}^H}\Delta {{\mathbf{z}}^*} = {{\mathbf{z}}^T}({{\mathbf{1}}^T} \otimes {\mathbf{I}})({\mathbf{E}} \odot {\mathbf{f}}{{\mathbf{f}}^H})({\mathbf{1}} \otimes {\mathbf{I}})\Delta {{\mathbf{z}}^*} ~\text{and}
\end{align}
\begin{align}
 \!\! {{\mathbf{z}}^T}{{\mathbf{F}}^T}{{\mathbf{F}}^*}\Delta {{\mathbf{z}}^*}& \!\!= \! {{\mathbf{z}}^T}({{\mathbf{1}}^T} \!\otimes\! {\mathbf{I}})({\mathbf{E}} \!\odot\! {\operatorname{vec} }({{\mathbf{F}}^T}){\operatorname{vec} }{({{\mathbf{F}}^T})^H})({\mathbf{1}} \! \otimes \! {\mathbf{I}})\Delta {{\mathbf{z}^*}} \nonumber \\
  & \!\!\!\!\!= {{\mathbf{z}}^T}({{\mathbf{1}}^T} \otimes {\mathbf{I}})({\mathbf{E}} \odot {\mathbf{Pf}}{{\mathbf{f}}^H}{{\mathbf{P}}^T})({\mathbf{1}} \otimes {\mathbf{I}})\Delta {{\mathbf{z}}^*},
\end{align}
where $\mathbf{P}$ is the permutation matrix and  ${\operatorname{vec} }({{\mathbf{F}}^T}) = {\mathbf{P}}{\operatorname{vec} }({\mathbf{F}}) = {\mathbf{Pf}}$.}

In what follows, we  first simplify the constraints in problem $\mathcal{P}4.1$ with \textbf{Lemma 1}. For constraint (82b), let  ${\mathbf{\tilde F}} = {\mathbf{f}}{{\mathbf{f}}^H}$, and ${\mathbf{\tilde W}} = {\mathbf{w}}{{\mathbf{w}}^H}$, we have
\begin{align}
{\eta _1}  = &\Delta {{\mathbf{h}}^H}{{\mathbf{A}}_s}{\mathbf{\tilde W}}{\mathbf{A}}_s^H\Delta {\mathbf{h}} + 2\operatorname{Re} \{ {\mathbf{\tilde h}}_s^H{\mathbf{\tilde W}}{\mathbf{A}}_s^H\Delta {\mathbf{h}}\} \nonumber\\
 &+ {\mathbf{\tilde h}}_s^H{\mathbf{\tilde W\tilde h}}_s^{},
\end{align}
where  $\Delta {\mathbf{h}} = \left[ \begin{gathered}
  \Delta {\mathbf{h}}_p^{} \hfill \\
  \Delta {\mathbf{h}}_s^{} \hfill \\
\end{gathered}  \right]$, ${{\mathbf{A}}_p} = \left[ \begin{gathered}
  {\mathbf{I}} \hfill \\
  {\mathbf{0}} \hfill \\
\end{gathered}  \right]$, ${{\mathbf{A}}_s} = \left[ \begin{gathered}
  {\mathbf{0}} \hfill \\
  {\mathbf{I}} \hfill \\
\end{gathered}  \right]$ and thus $\Delta {\mathbf{h}}_p^H = \Delta {{\mathbf{h}}^H}{{\mathbf{A}}_p}$, $\Delta {\mathbf{h}}_s^H = \Delta {{\mathbf{h}}^H}{{\mathbf{A}}_s}$.

Note that $\eta_3$ involves terms of both  ${\mathbf{\tilde h}}_s^H{\mathbf{F\tilde h}}_p^{}$ and $\Delta {\mathbf{h}}_s^H{\mathbf{F}}\Delta {\mathbf{h}}_p^{}$. However, it is difficult to tackle the product of these two terms, which is the second order of channel uncertainties. Hence, the ZF scheme is used to force the former term to zero, i.e.,  ${\mathbf{\tilde h}}_s^H{\mathbf{F}} = {\mathbf{0}}$, which is equivalent to
\begin{equation}
{\text{Tr}}\left( {({\mathbf{I}} \otimes {{{\mathbf{\tilde h}}}_s}{\mathbf{\tilde h}}_s^H){\mathbf{\tilde F}}} \right) = 0.
\end{equation}
Then, using \textbf{Lemma 1}, we have
\begin{equation}
{\eta _2} + {\eta _3} + {\eta _4} = \Delta {{\mathbf{h}}^H}{{\mathbf{{ K}}}_1}\Delta {{\mathbf{h}}} + 2\operatorname{Re} \{ {\mathbf{k}}_2^H\Delta {\mathbf{h}}\}  + {k_s},
\end{equation}
where
\begin{align}
&{{\mathbf{{ K}}}_1}\! = \!\rho {P_{SR}}{{\mathbf{A}}_s}({{\mathbf{\Phi }}_1} + {{\mathbf{\Phi }}_2}){\mathbf{\tilde F}}{({{\mathbf{\Phi }}_1} \!+\! {{\mathbf{\Phi }}_2})^H}{\mathbf{A}}_s^H + \rho {P_{PR}}({{\mathbf{A}}_s}{{\mathbf{\Phi }}_3} \nonumber \\
& \!\!+\! {{\mathbf{A}}_p}{{\mathbf{\Phi }}_2}){\mathbf{\tilde F}}{({{\mathbf{A}}_s}{{\mathbf{\Phi }}_3} \!+\! {{\mathbf{A}}_p}{{\mathbf{\Phi }}_2})^H} \!\!+\! {{\mathbf{A}}_s}{{\mathbf{\Psi }}_1}{\mathbf{A}}_s^H\!\! +\! {{\mathbf{A}}_s}{{\mathbf{\Psi }}_2}{\mathbf{A}}_s^H,
\end{align}
\begin{equation}
{\mathbf{k}}_2^H = {\mathbf{\tilde h}}_s^H{{\mathbf{\Psi }}_1}{\mathbf{A}}_s^H + {\mathbf{\tilde h}}_s^H{{\mathbf{\Psi }}_2}{\mathbf{A}}_s^H,
\end{equation}
\begin{equation}
{k_s} = {\mathbf{\tilde h}}_s^H({{\mathbf{\Psi }}_1} + {{\mathbf{\Psi }}_2}){\mathbf{\tilde h}}_s^{} + \sigma _s^2,
\end{equation}
\begin{equation}
{{\mathbf{\Phi }}_1} = {\mathbf{\tilde h}}_s^T \otimes {\mathbf{I}}, {{\mathbf{\Phi }}_2} = {\mathbf{I}} \otimes {\mathbf{\tilde h}}_s^T, {{\mathbf{\Phi }}_3} = {\mathbf{\tilde h}}_p^T \otimes {\mathbf{I}},
\end{equation}
\begin{equation}
  {{\mathbf{\Psi }}_1} = \rho {P_{PT}}({{\mathbf{g}}^T} \otimes {\mathbf{I}}){\mathbf{\tilde F}}{({{\mathbf{g}}^T} \otimes {\mathbf{I}})^H},
 \end{equation}
 \begin{equation}
  {{\mathbf{\Psi }}_2} = (\rho \sigma _r^2 + \sigma _c^2)({{\mathbf{1}}^T} \otimes {\mathbf{I}})({\mathbf{E}} \odot {\mathbf{\tilde F}})({\mathbf{1}} \otimes {\mathbf{I}}).
  \end{equation}
In (89), terms including the third or higher order of channel uncertainties are ignored due to their small values.

Substituting (87) and (89) into (82b), we can obtain
\begin{align}
&\Delta {{\mathbf{h}}^H}({{\mathbf{A}}_s}{\mathbf{\tilde W}}{\mathbf{A}}_s^H - t{{\mathbf{{ K}}}_1})\Delta {\mathbf{h}} + 2\operatorname{Re} \{ ({\mathbf{\tilde h}}_s^H{\mathbf{\tilde W}}{\mathbf{A}}_s^H - t{\mathbf{k}}_2^H)\Delta {\mathbf{h}}\} \nonumber\\
 &+ ({\mathbf{\tilde h}}_s^H{\mathbf{\tilde W\tilde h}}_s^{} - t{k_s}) \geq 0
\end{align}
So far, the first constraint (82b) is reformulated as (88) and (96).

In the similar way, the second constraint (82c) can be rewritten as
\begin{equation}
{\text{Tr}}\left( {({\mathbf{I}} \otimes {{{\mathbf{\tilde h}}}_p}{\mathbf{\tilde h}}_p^H){\mathbf{\tilde F}}} \right) = 0 ~ \text{and}
\end{equation}
\begin{align}
&\Delta {{\mathbf{h}}^H}({{\mathbf{A}}_p}{{\mathbf{\Psi }}_1}{\mathbf{A}}_p^H - \Gamma _p^{\min }{{\mathbf{{ K}}}_3})\Delta {\mathbf{h}} + 2\operatorname{Re} \{ ({\mathbf{\tilde h}}_p^H{{\mathbf{\Psi }}_1}{\mathbf{A}}_p^H -  \nonumber\\
&\Gamma _p^{\min }{\mathbf{k}}_4^H)\Delta {\mathbf{h}}\}  + ({\mathbf{\tilde h}}_p^H{{\mathbf{\Psi }}_1}{\mathbf{\tilde h}}_p^{} - \Gamma _p^{\min }{k_p}) \geq 0,
\end{align}
where
\begin{align}
&\!{{\mathbf{{ K}}}_3}\! =\! \rho {P_{PR}}{{\mathbf{A}}_p}({{\mathbf{\Phi }}_3} + {{\mathbf{\Phi }}_4}){\mathbf{\tilde F}}{({{\mathbf{\Phi }}_3} \!+\! {{\mathbf{\Phi }}_4})^H}{\mathbf{A}}_p^H \!+\! \rho {P_{SR}}({{\mathbf{A}}_p}{{\mathbf{\Phi }}_1} \nonumber\\
&\!\! +\! {{\mathbf{A}}_s}{{\mathbf{\Phi }}_4}){\mathbf{\tilde F}}({({{\mathbf{A}}_p}{{\mathbf{\Phi }}_1}\!\! +\! {{\mathbf{A}}_s}{{\mathbf{\Phi }}_4})^H}
  \!\! +\!  {{\mathbf{A}}_p}\!{\mathbf{\tilde W}}{\mathbf{\tilde A}}_p^H\!\! + \! {{\mathbf{A}}_p}\!{{\mathbf{\Psi }}_2}{\mathbf{A}}_p^H, \!
\end{align}
\begin{equation}
    {\mathbf{k}}_4^H =  {\mathbf{\tilde h}}_p^H{\mathbf{\tilde W}}{\mathbf{A}}_p^H + {\mathbf{\tilde h}}_p^H{{\mathbf{\Psi }}_2}{\mathbf{A}}_p^H,
\end{equation}
\begin{equation}
     {k_p} = {\mathbf{\tilde h}}_p^H( {\mathbf{\tilde W}} + {{\mathbf{\Psi }}_2}){\mathbf{\tilde h}}_p^{} + \sigma _p^2,
\end{equation}
\begin{equation}
  {{\mathbf{\Phi }}_4} = ({\mathbf{I}} \otimes {\mathbf{\tilde h}}_p^T).
\end{equation}

The third constraint (82d) can be equivalently converted as
\begin{equation}
  \Delta {{\mathbf{h}}^H}{{\mathbf{{ K}}}_5}\Delta {\mathbf{h}} + 2\operatorname{Re} \{ {\mathbf{k}}_6^H\Delta {\mathbf{h}}\}  + k \leq 0,
\end{equation}
where
\begin{equation}
{{\mathbf{{ K}}}_5} = {P_{PR}}{{\mathbf{A}}_p}{{\mathbf{\Psi }}_4}{\mathbf{A}}_p^H + {P_{SR}}{{\mathbf{A}}_s}{{\mathbf{\Psi }}_4}{\mathbf{A}}_s^H,
 \end{equation}
 \begin{equation}
 {\mathbf{k}}_6^H = {P_{PR}}{\mathbf{\tilde h}}_p^H{{\mathbf{\Psi }}_4}{\mathbf{A}}_p^H + {P_{SR}}{\mathbf{\tilde h}}_s^H{{\mathbf{\Psi }}_4}{\mathbf{A}}_s^H,
 \end{equation}
 \begin{align}
k = &{P_{PR}}{\mathbf{\tilde h}}_p^H{{\mathbf{\Psi }}_4}{\mathbf{\tilde h}}_p^{} + {P_{SR}}{\mathbf{\tilde h}}_s^H{{\mathbf{\Psi }}_4}{\mathbf{\tilde h}}_p^{} + {P_{PT}}{{\mathbf{g}}^H}{{\mathbf{\Psi }}_4}{\mathbf{g}} \nonumber \\
  &+ (\rho \sigma _r^2 + \sigma _c^2)\operatorname{Tr} ({\mathbf{\tilde F}})
   + \operatorname{Tr} ({\mathbf{\tilde W}}) - \xi \left( {1 - \rho } \right)\sigma _r^2,
    \end{align}
    \begin{equation}
    {{\mathbf{\Psi }}_3} = ({{\mathbf{1}}^T} \otimes {\mathbf{I}})({\mathbf{E}} \odot {\mathbf{P\tilde F}}{{\mathbf{P}}^T})({\mathbf{1}} \otimes {\mathbf{I}}),
    \end{equation}
    \begin{equation}
  {{\mathbf{\Psi }}_4} = \rho {{\mathbf{\Psi }}_3} - \xi (1 - \rho ){\mathbf{I}}.
  \end{equation}

  Next, we rely on the S-Procedure to further transform the re-expressions of constraints (82b)-(82d) into their corresponding tractable linear matrix inequality (LMI) forms.

\textit{\textbf{Lemma 2} (S-procedure \cite{boyd2004convex}):
Given Hermitian matrices  ${{\mathbf{\tilde A}}_i} \in {\mathbb{C}^{N \times N}}$ and  ${{\mathbf{\tilde b}}_i} \in {\mathbb{C}^{N \times 1}}$, ${\tilde c_i} \in \mathbb{R}, ~i = 1,2,3.$ Define the functions  ${f_i}({\mathbf{x}}) = {{\mathbf{x}}^H}{{\mathbf{\tilde A}}_i}{\mathbf{x}} + 2\operatorname{Re} \{ {\mathbf{\tilde b}}_i^H{\mathbf{x}}\}  + {\tilde c_i}$. Then,  ${f_1}({\mathbf{ x}}) \geq 0$ and ${f_2}({\mathbf{ x}}) \geq 0$
imply  ${f_3}({\mathbf{ x}}) \geq 0$, if and only if there exist  ${\varsigma _1} \geq 0$ and ${\varsigma _2} \geq 0$ such that
\begin{equation}
\left[ \begin{gathered}
  {{{\mathbf{\tilde A}}}_3}{\text{ ~ }}{{{\mathbf{\tilde b}}}_3} \hfill \\
  {\mathbf{\tilde b}}_3^H{\text{~~   }}{c_3} \hfill \\
\end{gathered}  \right] - {\varsigma _2}\left[ \begin{gathered}
  {{{\mathbf{\tilde A}}}_2}{\text{~~\! }}{{{\mathbf{\tilde b}}}_2} \hfill \\
  {\mathbf{\tilde b}}_2^H{\text{  ~ }}{c_2} \hfill \\
\end{gathered}  \right] - {\varsigma _1}\left[ \begin{gathered}
  {{{\mathbf{\tilde A}}}_1}{\text{ ~ }}{{{\mathbf{\tilde b}}}_1} \hfill \\
  {\mathbf{\tilde b}}_1^H{\text{ ~  }}{c_1} \hfill \\
\end{gathered}  \right] \succeq   {\mathbf{0}}
\end{equation}
provided that there exists a vector $\tilde {\mathbf{x}}$ with ${f_1}({\mathbf{\tilde x}}) > 0$ and ${f_2}({\mathbf{\tilde x}}) > 0$.}

In this paper, we can take
\begin{equation}
{f_1}(\Delta {\mathbf{h}}) =  - \Delta {{\mathbf{h}}^H}{{\mathbf{A}}_p}{\mathbf{A}}_p^H\Delta {\mathbf{h}} + \varepsilon _p^2 \geq 0 ~\text{and}
\end{equation}
\begin{equation}
{f_2}(\Delta {\mathbf{h}}) =  - \Delta {{\mathbf{h}}^H}{{\mathbf{A}}_s}{\mathbf{A}}_s^H\Delta {\mathbf{h}} + \varepsilon _s^2 \geq 0,
\end{equation}
For the first constraint (82b), by employing \textbf{lemma 2}, the LMI form of (96) is given as
\begin{equation}
\left[ \begin{gathered}
   \mathbf{T}_s+{\varsigma _{s1}}{{\mathbf{A}}_p}{\mathbf{A}}_p^H+{\varsigma _{s2}}{{\mathbf{A}}_s}{\mathbf{A}}_s^H{\text{~~ ~~           }}{\mathbf{A}}_s^{}{\mathbf{\tilde W\tilde h}}_s - t{\mathbf{k}}_2^{} \hfill \\
  {\text{    ~~~ ~        }}{\mathbf{\tilde h}}_s^H{\mathbf{\tilde W}}{\mathbf{A}}_s^H - t{\mathbf{k}}_2^H{\text{           ~~~~~~~~~            }} t_s- {\varsigma _{s1}}\varepsilon _p^2 - {\varsigma _{s2}}\varepsilon _s^2 \hfill \\
\end{gathered}  \right] \succeq   {\mathbf{0}},
\end{equation}
where $\mathbf{T}_s={{\mathbf{A}}_s}{\mathbf{\tilde W}}{\mathbf{A}}_s^H- t{{\mathbf{{ K}}}_1}$, $t_s={\mathbf{\tilde h}}_s^H{\mathbf{\tilde W\tilde h}}_s^{} - t{k_s}$ and ${\varsigma _{s1}}$, ${\varsigma _{s2}}$ are introduced variables.

Similarly, for the secondary constraint (82c), (98) can be transformed as
\begin{equation}
\left[ \begin{gathered}
 \mathbf{T}_p +{\varsigma _{p1}}{{\mathbf{A}}_p}{\mathbf{A}}_p^H+{\varsigma _{p2}}{{\mathbf{A}}_s}{\mathbf{A}}_s^H{\text{     ~~        }}{\mathbf{A}}_p^{}{\mathbf{\Psi }}_1{\mathbf{\tilde h}}_p^{} - \Gamma _p^{\min }{\mathbf{k}}_4^{} \hfill \\
  {\text{     ~~   }}{\mathbf{\tilde h}}_p^H{{\mathbf{\Psi }}_1}{\mathbf{A}}_p^H - \Gamma _p^{\min }{\mathbf{k}}_4^H{\text{                   ~~~~~~  ~       }}t_p - {\varsigma _{p1}}\varepsilon _p^2 - {\varsigma _{p2}}\varepsilon _s^2 \hfill \\
\end{gathered}  \right] \succeq   {\mathbf{0}},
\end{equation}
where  $\mathbf{T}_p={{\mathbf{A}}_p}{{\mathbf{\Psi }}_1}{\mathbf{A}}_p^H - \Gamma _p^{\min }{{\mathbf{{ K}}}_3}$,  $t_p={\mathbf{\tilde h}}_p^H{{\mathbf{\Psi }}_1}{\mathbf{\tilde h}}_p^{} - \Gamma _p^{\min }{k_p}$  and ${\varsigma _{p1}}$, ${\varsigma _{p2}}$ are introduced variables.

And for the third constraint (82d), (103) can be rewritten as
\begin{equation}
\left[ \begin{gathered}
   - {{\mathbf{{ K}}}_5}+{\nu _1}{{\mathbf{A}}_p}{\mathbf{A}}_p^H+{\nu _2}{{\mathbf{A}}_s}{\mathbf{A}}_s^H{\text{  ~~~  ~~~      }} - {{\mathbf{k}}_6} \hfill \\
  {\text{     ~~~~~~~~~     }} - {\mathbf{k}}_6^H{\text{          ~ ~~~~~~~~~~~~~~             }} - k - {\nu _1}\varepsilon _p^2 - {\nu _2}\varepsilon _s^2 \hfill \\
\end{gathered}  \right] \succeq {\mathbf{0}},
\end{equation}
where ${\nu _1}$ and ${\nu _1}$ are introduced variables.

Therefore, with given $t$ and $\rho$, problem $\mathcal{P}4.1$ can be re-expressed as a feasible SDR problem ($\mathcal{P}4.2$)
\begin{subequations}
\begin{align}
&\underset{{{\tilde {\mathbf{F}}}, \tilde {{\mathbf{W}}} }}{\text{max}}&& ~~0\\
&\text{s. t.}&&  (88), (97),  (112)-(114),  \\
&&&\!\!\!\!\!\!\!\!\!\!\!\!\!\!\!\!\!\!\!\!\!\!\!\!{\varsigma _{s1}} \geq 0, {\varsigma _{s2}} \geq 0,{\varsigma _{p1}} \geq 0,{\varsigma _{p2}} \geq 0,{\nu _1} \geq 0,{\nu _2} \geq 0,\\
&&& ~~~~{{\tilde {\mathbf{F}}}\succeq  {\mathbf{0}},\tilde {{\mathbf{W}}}\succeq  {\mathbf{0}}},
\end{align}
\end{subequations}
which is  convex and can be effectively solved by off-the-shelf solvers, such as CVX \cite{grantmay}.
The optimal $t$ can be found via bisection and the optimal $\rho$  can be achieved via the exhaustive search. Hence, the algorithm 2 for the robust scheme is listed as below. What is noteworthy is that the Gaussian randomization method \cite{luo2010semidefinite} can be employed to extract the rank-one solution if the rank of obtained solution is greater than one.

\begin{algorithm}[h]
    \caption{ Robust solution to problem $\mathcal{P}$4 with imperfect CSI }
    \begin{algorithmic}[1]
    \STATE {Initialize a step size $\rho_s$ for $\rho$;}
    \FOR {$\rho=0:\rho_s:1$}
    \STATE {Initialize $t^{\min}$, $t^{\max}$ and tolerance $\delta_t$;}
     \WHILE {$ { t^{\max}-t^{\min} > \delta_t } $}
    \STATE $t \leftarrow (t^{\min}+t^{\max})/2$;
    \STATE Check the feasibility of problem $\mathcal{P}$4.2 to via CVX;
    \IF{It is feasible}
    \STATE $t^{\min} \leftarrow t $;
    \ELSE
    \STATE $t^{\max} \leftarrow t $;
    \ENDIF
    \ENDWHILE
    \ENDFOR
    \STATE Find the maximum $t$ as $t^*$ and its related $\rho^*$;
    \RETURN $ \tilde {\mathbf{F}}^*$, $\tilde{\mathbf{ W}}^*$,  $\rho^*$ and $t^*$;
    \label{code:recentEnd}
    \end{algorithmic}
\end{algorithm}

\section{Simulation Results}

In this section, we evaluate the performance of our proposed DWPT scheme via computer numerical simulations.  For simplicity, the received noise power is $\sigma_r^2=\sigma_c^2=\sigma_p^2=\sigma _s^2=1$ mW. Unless otherwise specified, other simulation parameters are set as follows. Assume that  ST has  $M=4$ antennas and the transmission power at PT, PR and SR is $P_{PT}=30$ dBm,  $P_{PR}=P_{SR}=30$ dBm. The energy conversion efficiency is  $\xi=50\%$ and the minimal rate requirement of PU  is $R_p^{min}=2.5$ bps/Hz.  As described in Fig. 2, we consider a simple scenario where locations of PT, ST, PR and SR are (-5, 0), (0, 0), (5, -1) and (5, 1) in coordinates, respectively. The distance unit is in meters and the path-loss exponent is 2. All channel entries are independently generated from i.i.d Rayleigh fading with their respective average power values. For comparison, the energy harvesting cognitive radio system without  power transfer from destinations (i.e., PR and SR) \cite{zheng2014information} is also considered, which is labeled as `w/o destination-aided'. Except for Fig. 3, the simulation results  are achieved over 500 independent channel realizations.
\begin{figure}
\centering
\includegraphics[width=9.0cm, height=4.1cm]{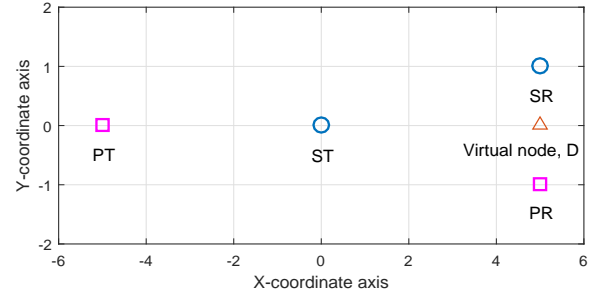}
\caption{Locations of PT, ST, PR and SR in X-Y coordinations.}
\label{systemmodel}
\end{figure}

\subsection{Performance Evaluation for the Perfect CSI case}
\begin{figure}
\centering
\includegraphics[width=9.0cm, height=7.0cm]{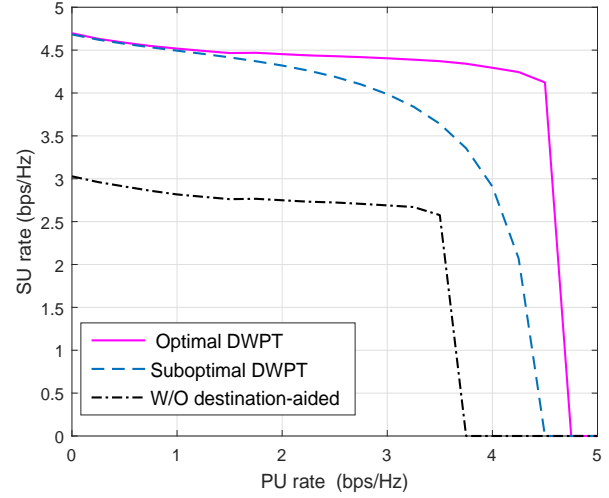}
\caption{SU-PU rate regions with $P_{PR}=P_{SR}=30$ dBm.}
\label{systemmodel}
\end{figure}

At first,  the achievable SU-PU rate regions   are characterized in Fig. 3 for different schemes.  A specific channel realization is randomly chosen as $\mathbf{g}=[-0.2694-0.0123i, -0.2221-0.0584i, -0.1695+0.2270i, -0.1823+0.1044i]^T$, $\mathbf{h}_p=[-0.0762-0.1064i, 0.0060+0.2268i, 0.0962-0.3864i, 0.0037+0.0652i]^T$ and $\mathbf{h}_s=[-0.0036+0.0617i, -0.1718-0.0510i, 0.0218-0.1389i, 0.0480+0.2174i]^T$.
  It is observed that both optimal and suboptimal  DWPT schemes achieve significantly larger rate region than   the  `w/o destination-aided' scheme due to the destinations' energy transfer. In addition, the optimal DWPT scheme always outperforms the suboptimal DWPT scheme. This is owing to the fact that the spatial degrees of freedom for the suboptimal scheme are slightly reduced by the decomposition of relay matrix $\mathbf{F}$ and the ZF design of receiver filter $\mathbf{f}_r$.
Moreover, the SU rate  of the low-complexity suboptimal DWPT scheme closely approaches to that of the optimal DWPT scheme when $R_p^{min} \leq 1.5$ bps/Hz. This is because that when the value of $R_p^{min}$ is small, the allocated power for relay matrix $\mathbf{F}$ is extremely low such that the suboptimal design of $\mathbf{F}$ has little effect on the SU rate.

\begin{figure}
\centering
\includegraphics[width=9.0cm, height=7.0cm]{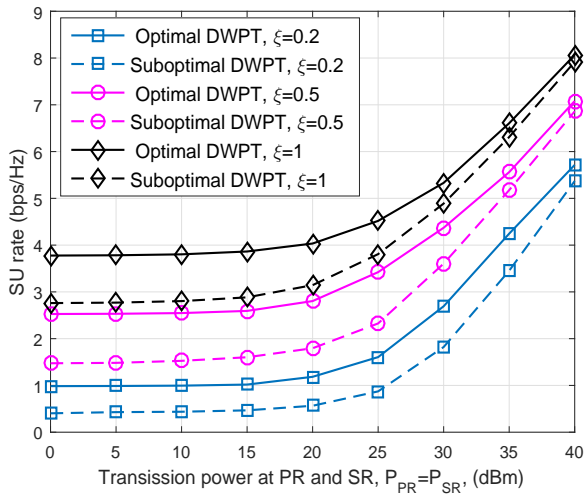}
\caption{Achievable SU rate versus transmission power at destinations (PR and SR)  with  $R_p^{min}=2.5$ bps/Hz.}
\label{systemmodel}
\end{figure}

In Fig. 4,  the  impact of transmission power at PR and SR on the achievable SU rate is investigated with different energy conversion efficiencies $\xi$. We assume that transmission power at PR and SR is equal to each other in our simulations. It is straightforward that the SU rate is improved  as the transmission power at destinations increases.  Note that, on one hand, the power aided by destinations enhances the amount of  harvested energy at ST in the first phase from (3). On the other hand, based on (7) and (8), this power also brings additional interferences to PR  and SR in the second phase. The continuous increasing SU rate with respect to transmission power at PR and SR indicates that, the interferences caused by destinations' power transfer can be well suppressed and the desired signals  with dominant power can benefit a lot from our proposed DWPT scheme. Besides, we can observe that the optimal DWPT scheme has obvious performance gain over the suboptimal scheme for different energy conversion efficiencies $\xi$, while the gap between them gradually reduces when the transmission power at destinations increases. The reason is  that when the transmission power at destinations is high, the system becomes interference-limited \cite{tse2005fundamentals}, and thus the suboptimal scheme with  ZF-based receiver filter $\mathbf{f}_r$ can cancel the strong interferences and approximately achieve the optimal performance.

\begin{figure}
\centering
\includegraphics[width=9.0cm, height=7.0cm]{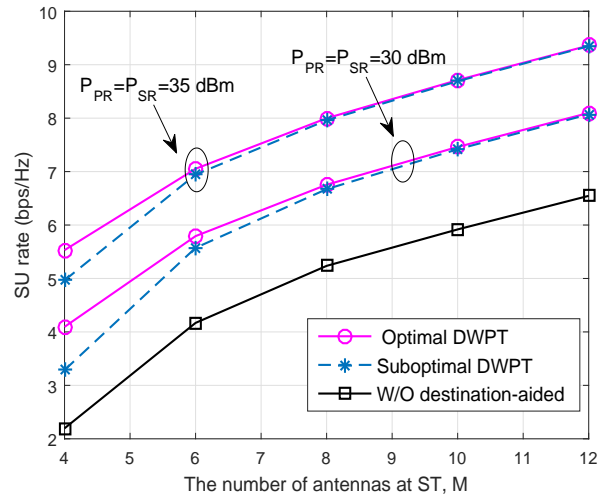}
\caption{Achievable SU rate versus the number of antennas at M with  $R_p^{min}=2.5$ bps/Hz.}
\label{systemmodel}
\end{figure}

Fig. 5 plots the achievable SU rates versus the number of antennas at ST for different schemes with  $R_p^{min}=2.5$ bps/Hz. For all schemes, as expected, more antennas employed by ST will result in better SU rate performance. In addition, it can be easily found that, the SU rate performance of the suboptimal scheme is gradually approaching to that of the optimal scheme with the increasing number of antennas at ST, especially  when $M\geq8$. This is mainly because  that, for the suboptimal scheme,  the spatial degrees of freedom loss caused by the ZF-based receiver filter $\mathbf{f}_r$  can be improved as $M$ increases.

\begin{figure}
\centering
\includegraphics[width=9.0cm, height=7.0cm]{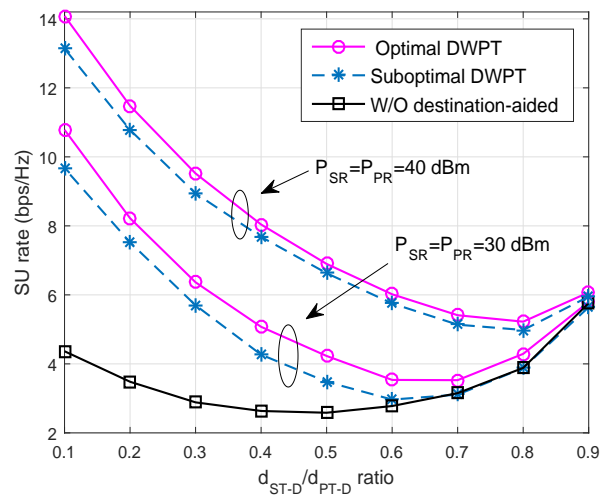}
\caption{Achievable SU rate versus $d_{ST-D}/d_{PT-D}$ with $R_p^{min}=2.5$ bps/Hz.}
\label{systemmodel}
\end{figure}

To illustrate the impact of ST's location on the SU rate in Fig. 6, we assume that ST can move along the X-coordinate axis from PT to D, a virtual node  situated at (5, 0) as plotted in Fig. 2.  $d_{ST-D}$ and $d_{PT-D}$ respectively denote distances between ST and D, PT and D.
From Fig. 6, it  can be easily observed that our proposed DWPT scheme is greatly preferred  when ST is close to PR and SR, since the amount of harvested energy at ST is effectively enhanced. Besides, it is of interest to find that, with the ST's movement from PT to D (i.e., $d_{ST-D}/d_{PT-D}$ from 0.9 to 0.1), the achieved SU rates for all three schemes first decrease and then increase. And the worst point is moving to PT when transmission power of PR and SR is increasing. More curiously, the  worst point of the `w/o destination-aided' scheme occurs when $d_{ST-D}/d_{PT-D}$ ratio is 0.5  rather than 0.1. This is mainly due to the fact that, the harvested energy at ST is not enough to offset the severe path-loss between ST and PR/SR when ST is in the middle location.

\begin{figure}
\centering
\includegraphics[width=9.0cm, height=7.0cm]{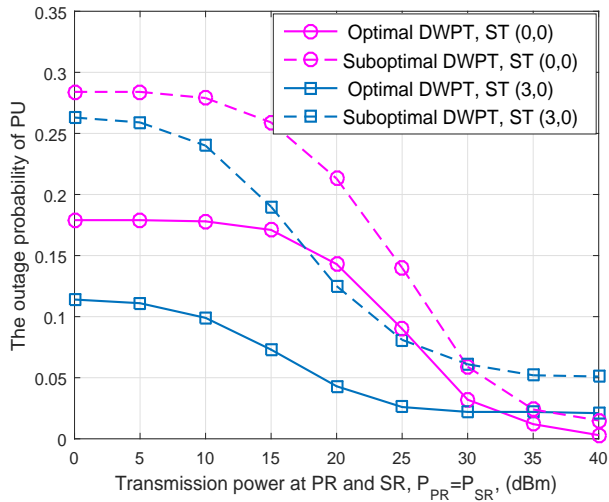}
\caption{The outage performance of  PU  versus transmission power at PR and SR with $R_p^{min}=2.5$ bps/Hz.}
\label{systemmodel}
\end{figure}

In addition to the performance evaluation of achievable SU rate, we also investigate the outage performance of PU in Fig. 7. The outage will occur when the required rate demand of PU cannot be guaranteed. That is, the considered problem is infeasible.
 `ST (0, 0)' and `ST (3, 0)' in Fig. 7 respectively mean that ST is situated at (0, 0) and (3, 0) in Fig. 2. From Fig. 7, we can observe that the outage probability of PU is declining  with the increase of transmission power at PR and SR. Combining Fig. 4 and Fig. 7, it is noted that when ST is located at (0, 0) and $P_{PR}=P_{SR}\geq 30$ dBm, the growth trend of SU rate is very evident and the outage probability of PU is close to zero.  This reveals that, not only SU but also PU can benefit from our proposed DWPT scheme.
Furthermore, the proposed optimal DWPT scheme achieves better outage performance than the suboptimal scheme, and the gap between them reduces as transmission power at PR and SR increases. The reason behind this phenomenon is similar to Fig. 4.
Besides, it can also be found that the outage performance of `ST (0, 0)'  unexpectedly outperforms that of `ST (3, 0)' for both optimal and suboptimal DWPT schemes when $P_{PR}=P_{SR}\geq30$ dBm. This is mainly because that the interference at PR is stronger when ST is located at (3, 0). However, this slight worse outage performance of PU does not affect the improvement of SU rate when ST moves from (0, 0) to (3, 0) (i.e., $d_{ST-D}/d_{PT-D}$ from 0.5 to 0.2)  according to  Fig. 6.

\subsection{Performance Evaluation for the Imperfect CSI case}
\begin{figure}
\centering
\includegraphics[width=9.0cm, height=7.0cm]{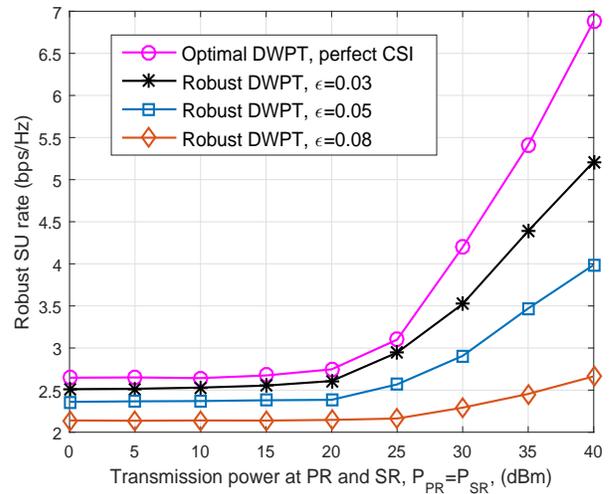}
\caption{Robust SU rate versus transmission power at PR and SR  with  $R_p^{min}=2.5$ bps/Hz.}
\label{systemmodel}
\end{figure}

This subsection shows the performance evaluation for the imperfect CSI case. For simulation, we assume that the radii of channel uncertainty regions for $\mathbf{h}_p$ and $\mathbf{h}_s$ are equal to each other, i.e., $\varepsilon_p=\varepsilon_s=\varepsilon$. Other simulation parameters are the same as the perfect CSI case.

   The achieved worst-case  SU rates versus transmission power at PR and SR are characterized in Fig. 8 for different levels of channel uncertainty.  The rate demand of PU is set as $R_p^{min}=2.5$ bps/Hz.  As can be seen, for both perfect and imperfect CSI cases, the larger value of transmission power at destinations, the better rate performance SU has. Furthermore, Observing from this figure, we can see that with the increase of channel uncertainty level, the achieved performance in terms of the worst-case SU rate is deteriorated.

\begin{figure}
\centering
\includegraphics[width=9.0cm, height=7.0cm]{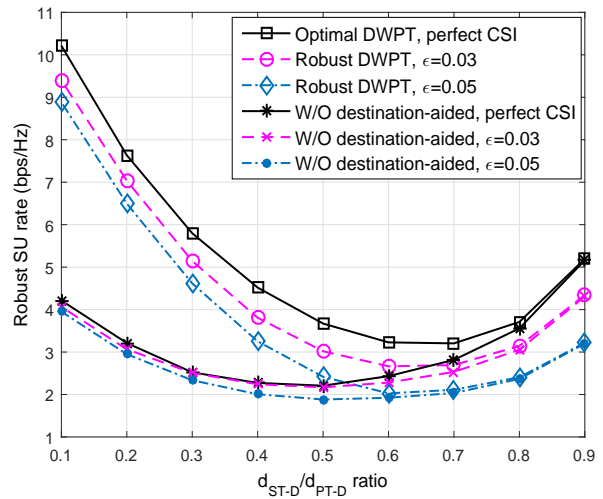}
\caption{Robust SU rate versus $d_{ST-D}/d_{PT-D}$ for different schemes  with $P_{PR}=P_{SR}=30$ dBm and $R_p^{min}=2.5$ bps/Hz.}
\label{systemmodel}
\end{figure}

In Fig. 9, the impact of ST's location on the worst-case robust SU rate is presented for different schemes. We assume that the rate demand of PU is $R_p^{min}=2.5$ bps/Hz and the transmission power at each nodes is $P_{PR}=P_{SR}=30$ dBm. Similar to  the perfect CSI case, we can observe that our proposed robust DWPT scheme is also more preferred for all different channel uncertainty levels when ST is closer to destinations. Nevertheless, note that when ST is much closer to PT (i.e., $d_{ST-D}/d_{PT-D} \to 0.9$),  the  SU rates  of the DWPT scheme for both perfect and imperfect CSI cases will fall to that of the `w/o destination-aided' scheme. In this case, it is not necessary for destinations to assist power to ST. 

\section{Conclusion}

This paper has proposed a DWPT scheme for a cognitive relay network, where the multiple-antenna energy-limited ST  first harvests the energy sent by PT as well as PR and SR, and then relays the traffic from PT to PR and also serves SR. The relay process matrix, cognitive beamforming vector and power splitter have been jointly optimized to maximize the SU rate with the energy causality constraint and the constraint that the  rate requirement of PU is met. Both the perfect and imperfect CSI scenarios have been investigated. For the former case, the global optimal and low-complexity suboptimal solutions have been presented. For the latter case, a worst-case robust algorithm has been proposed.
 It has been demonstrated in the simulation that our proposed DWPT scheme is greatly preferred when ST is close to PR and SR. Thus, the location-based relay selection scheme could be our future work.

\appendices
\section{Proof of Proposition 1}
Without loss of generality, similar to \cite{zheng2013cooperative}, $\mathbf{F}$ can be expressed as
\begin{align}
{\mathbf{F}}& = \left[ {{{\mathbf{V}}_1}{\text{ }}{\mathbf{V}}_1^ \bot } \right]\left[ \begin{gathered}
  {\mathbf{A}}{\text{   }}{\mathbf{B}} \hfill \\
  {\mathbf{C}}{\text{   }}{\mathbf{D}} \hfill \\
\end{gathered}  \right]\left[ \begin{gathered}
  {\mathbf{U}}_2^H \hfill \\
  {\mathbf{U}}_2^{ \bot H} \hfill \\
\end{gathered}  \right]\\
 &= {{\mathbf{V}}_1}{\mathbf{AU}}_2^H{\text{ + }}{{\mathbf{V}}_1}{\mathbf{BU}}_2^{ \bot H} + {\mathbf{V}}_1^ \bot {\mathbf{CU}}_2^H + {\mathbf{V}}_1^ \bot {\mathbf{DU}}_2^{ \bot H} \nonumber
\end{align}
where  ${\mathbf{V}}_1^ \bot  \in {\mathbb{C}^{M \times (M - 2)}}$, ${\mathbf{U}}_2^ \bot  \in {\mathbb{C}^{M \times (M - 3)}}$, ${\mathbf{A}} \in {\mathbb{C}^{2 \times 3}}$, ${\mathbf{B}} \in {\mathbb{C}^{2 \times (M - 3)}}$, ${\mathbf{C}} \in {\mathbb{C}^{(M - 2) \times 3}}$ and ${\mathbf{D}} \in {\mathbb{C}^{(M - 2) \times (M - 3)}}$. Obviously,  ${\mathbf{V}}_1^H{\mathbf{V}}_1^ \bot  = {\mathbf{0}}$, ${\mathbf{U}}_2^{ \bot H}{\mathbf{U}}_2^{} = {\mathbf{0}}$. Thus, we have
\begin{equation}
\left[ \begin{gathered}
  {\mathbf{h}}_{\text{s}}^H \hfill \\
  {\mathbf{h}}_{\text{p}}^H \hfill \\
\end{gathered}  \right]{\mathbf{V}}_1^ \bot  = {\mathbf{0}}, {\mathbf{U}}_2^{ \bot H}\left[ {{{\mathbf{h}}_{\text{s}}},{\text{ }}{{\mathbf{h}}_{\text{p}}},{\text{ }}{\mathbf{g}}} \right]{\text{ = }}{\mathbf{0}}.
\end{equation}

 Substituting (116) into terms related to $\mathbf{F}$ in problem $\mathcal{P}1$, we know that $\mathbf{B}$, $\mathbf{C}$ and $\mathbf{D}$ do not affect  ${\left| {{\mathbf{h}}_s^H{\mathbf{Fg}}} \right|^2}$, ${\left| {{\mathbf{h}}_s^H{\mathbf{F}}{{\mathbf{h}}_p}} \right|^2}$ and ${\left| {{\mathbf{h}}_p^H{\mathbf{Fg}}} \right|^2}$. In addition, $\mathbf{C}$ and $\mathbf{D}$ have no impact on ${\left\| {{\mathbf{h}}_s^H{\mathbf{F}}} \right\|_2^2}$ and ${\left\| {{\mathbf{h}}_p^H{\mathbf{F}}} \right\|_2^2}$, $\mathbf{B}$ and $\mathbf{D}$ have no effect on $\left\| {{\mathbf{Fg}}} \right\|_2^2$, $\left\| {{\mathbf{F}}{{\mathbf{h}}_s}} \right\|_2^2$ and $\left\| {{\mathbf{F}}{{\mathbf{h}}_p}} \right\|_2^2$.
As a result, from (9) and (10), it is observed that $\mathbf{C}$ and $\mathbf{D}$ have no effect on  SINRs of PR and SR. However, from (6), they increase the transmission power at ST. Hence, the optimal choice of $\mathbf{C}$ and $\mathbf{D}$ is $\mathbf{C}=\mathbf{0}$ and $\mathbf{D}=\mathbf{0}$. Besides, both SINRs of PR and SR are increased if we set $\mathbf{B}=\mathbf{0}$. Therefore,  ${\mathbf{F}}{\text{ = }}{{\mathbf{V}}_1}{\mathbf{AU}}_2^H$.

Similarly, $\mathbf{w}$ can be expressed as
\begin{equation}{\mathbf{w}}{\text{ = }}\left[ {{{\mathbf{V}}_1},{\text{ }}{\mathbf{V}}_1^ \bot } \right]\left[ \begin{gathered}
  {\mathbf{b}} \hfill \\
  {\mathbf{c}} \hfill \\
\end{gathered}  \right] = {{\mathbf{V}}_1}{\mathbf{b}} + {\mathbf{V}}_1^ \bot {\mathbf{c}},
 \end{equation}
 where ${\mathbf{b}} \in {\mathbb{C}^{2 \times 1}}$ and ${\mathbf{c}} \in {\mathbb{C}^{(M - 2) \times 1}}$. Note that $\mathbf{c}$ has no impact on ${\left\| {{\mathbf{h}}_s^H{\mathbf{w}}} \right\|_2^2}$, ${\left\| {{\mathbf{h}}_p^H{\mathbf{w}}} \right\|_2^2}$,  and thus does not affect SINRs at PR and SR. But it increases the transmission power at ST. Hence,  $\mathbf{c}=\mathbf{0}$ and $\mathbf{w}=\mathbf{V}_1\mathbf{b}$. This completes the proof. \IEEEQED


%


\ifCLASSOPTIONcaptionsoff
  \newpage
\fi



%

\bibliographystyle{IEEEtran}

\bibliography{reference}
%

%
%
%




\end{document}